\documentclass[aps,prl,nofootinbib,longbibliography,superscriptaddress,twocolumn]{revtex4-2}
\usepackage{amsmath}
\usepackage{amssymb}
\usepackage{graphicx}
\usepackage{bm}
\usepackage[colorlinks=true,linkcolor=cyan,anchorcolor=cyan, citecolor=cyan,urlcolor=cyan]{hyperref}
\usepackage[mathlines]{lineno}
\usepackage{gensymb}

\begin{document}
\title{Relaxation and Its Effects on Electronic Structure in Twisted Systems: An Analytical Perspective}
\author{Junxi Yu}
\author{Bingbing Wang}
\author{Cheng-Cheng Liu}
\email{ccliu@bit.edu.cn}
\affiliation{Centre for Quantum Physics, Key Laboratory of Advanced Optoelectronic Quantum Architecture and Measurement (MOE), School of Physics, Beijing Institute of Technology, Beijing 100081, China}
\begin{abstract}
Lattice relaxation profoundly reshapes electronic structures in twisted materials. Prevailing treatments, however, typically rely on large-scale density functional theory (DFT), which is computationally costly and mechanistically opaque. Here, we develop a unified analytical framework to overcome these limitations. From continuum elastic theory, we derive closed-form solutions for both in-plane and out-of-plane relaxation fields. We further introduce an analytical phase factor expansion theory that maps relaxation into the electronic Hamiltonian. By applying this framework, the relaxation-mediated single-particle and many-body topological phase transitions in twisted MoTe$_{2}$ is accurately captured, and the evolution of ﬂat bands in magic-angle graphene is quantitatively reproduced. Our work transforms the research of moiré relaxation from black-box numerical fitting to an analytical paradigm, offering fundamental insights, exceptional efficiency, and general applicability to a wide range of twisted materials.
\end{abstract}
\maketitle
\textit{Introduction.---}
Twisting has become a fundamental tool for engineering quantum states in two-dimensional (2D) materials in recent years. Twisted systems host a series of remarkable quantum phenomena, including unconventional superconductivity, correlated insulating states, fractional topological states, and so on~\cite{cao_correlated_2018,serlin_intrinsic_2020,cao_unconventional_2018,yankowitz_tuning_2019,lu_superconductors_2019,cai_signatures_2023,zengThermodynamicEvidenceFractional2023,xuObservationIntegerFractional2023,guo2024superconductivity,hao_electric_2021,kang_evidence_2024,lu_fractional_2024,regan_mott_2020,xie_tunable_2025}.

Lattice relaxation plays a crucial role in the formation mechanisms of these exotic states~\cite{yamamoto_princess_2012,woods_commensurate_2014,jung_origin_2015,san-jose_electronic_2014,koshino_maximally_2018,koshino_effective_2020, nam_lattice_2017,mao_transfer_2024a,wang_fractional_2024,yu_fractional_2024,zhang_polarizationdriven_2024,liuPseudoLandaulevelRepresentationTwisted2019,tarnopolsky_origin_2019,xie_lattice_2023,yu_general_2025}. For instance, in twisted bilayer graphene (TBG), relaxation governs the opening of flat-band gaps and the evolution of topological properties. On the other hand, twisted MoTe$_{2}$ (tMoTe$_{2}$) has emerged as a new star system following the discovery of the fractional Chern insulator (FCI) state, where relaxation significantly influences topological phase transitions. Notably, a significant discrepancy exists in the topological phase transition boundary 
predicted by conventional rigid models and density 
functional theory (DFT) calculations. 

The prevailing approach to incorporating the influence of lattice relaxation into electronic structure involves fitting the band structures obtained from large-scale DFT calculations to parameterize continuum models~\cite{zhang_polarizationdriven_2024,morales-duran_magic_2024,wang_fractional_2024,yu_fractional_2024,reddy_fractional_2023,jia_moire_2024a,xu_maximally_2024}. However, this paradigm faces three major challenges: (1) Computational non-scalability. Large-scale (relativistic) DFT calculations involving tens of thousands of atoms are prohibitively expensive, and separate calculations are required for different parameters, making comprehensive parameter space scans impractical; (2) Model uncertainty. Band-structure fitting suffers from parameter uncertainty in a high-dimensional space: distinct parameter sets yield nearly indistinguishable bands, preventing a unique determination of the underlying physics; (3) Lack of physical insight. While DFT provides quantitative results as a black box, it fails to reveal the physical mechanisms by which relaxation affects electronic structure. These limitations severely hinder the development of twistronics. Thus, there is an urgent need for an effective framework to account for the lattice relaxation and its effect on electronic structures in twisted systems.

This work presents an analytical solution. We develop an analytical theory of relaxation fields, rigorously deriving a closed-form solution for complete relaxation starting from continuum elastic theory. Furthermore, we employ an analytical phase factor expansion theory, achieving a comprehensive treatment of in-plane relaxation effects on electronic structure through a series expansion. On the other hand, the coupling function is treated as a functional of the out-of-plane displacement field, quantifying its effect on the electronic structure. We construct an analytical framework to quantitatively characterize relaxation and its effects on the electronic structures in twisted systems. As representative applications, our analytical framework captures single-particle topological phase transitions in tMoTe$_{2}$ by reproducing the DFT-predicted transition point and characterizes the inﬂuence of relaxation on the FCI correlated phase. Furthermore, it successfully captures the relaxed flat bands evolution in magic-angle TBG. The established theoretical framework offers a computational speedup of $> 10^4$ times and uncovers the mechanism of relaxation-electron coupling.

\textit{Analytic Solution for Relaxation.---}
First, we define the displacement field induced by relaxation of the twisted system:
$\pmb{u}^{(l)}(\pmb{r})=\pmb{u}_{\parallel}^{(l)}(\pmb{r})+\pmb{u}_{\perp}^{(l)}(\pmb{r}),$
where we decompose it into in-plane relaxation $\pmb{u}_{\parallel}^{(l)}(\pmb{r})=u_{x}^{(l)}\hat{\pmb{e}}_{x}+u_{y}^{(l)}\hat{\pmb{e}}_{y}$ and out-of-plane relaxation $\pmb{u}_{\perp}^{(l)}(\pmb{r})=u_{z}^{(l)}\hat{\pmb{e}}_{z}$. Here, $l=t,b$ label the top layer and bottom layer,  and $\hat{\pmb{e}}_{x},\hat{\pmb{e}}_{y},\hat{\pmb{e}}_{z}$ denote unit vectors in the $x,y,z$ directions, respectively. Defining the transformation $\pmb{u}^{\pm}(\pmb{r})=\pmb{u}^{(t)}(\pmb{r})\pm \pmb{u}^{(b)}(\pmb{r})$, we construct the Lagrangian, yielding the Euler-Lagrange equations,
\begin{equation}\label{eq:E-L eqs}
    \frac{\partial L}{\partial u_{i}^{\pm}}-\sum_{j\in \{x,y,z\}}\partial_{j} \frac{\partial L}{\partial (\partial_{j} u_{i}^{\pm})}=0,
\end{equation}
where $i\in \{x,y,z\}$, comprising six equations. The solution depends on the specific form of the Lagrangian density. Take the 2D hexagonal Bravais lattice as an example, the static energy contribution to the Lagrangian density reads~\cite{landau1986}
\begin{equation}\label{eq:LE}
    \begin{aligned}
        L_{E} = 
    \sum_{\pm}
    \{
    &\frac{\mu}{4}\left(\frac{\partial u_{x}^{\pm}}{\partial x}-\frac{\partial u_{y}^{\pm}}{\partial y}\right)^{2}+\frac{\mu}{4}\left(\frac{\partial u_{x}^{\pm}}{\partial y}    +\frac{\partial u_{y}^{\pm}}{\partial x}\right)^{2}
    \\
    &+\frac{\zeta}{2}\left[\left(\frac{\partial u_{z}^{\pm}}{\partial x}\right)^{2}+\left(\frac{\partial u_{z}^{\pm}}{\partial y}\right)^{2}\right]
    \}.
    \end{aligned}
\end{equation}
Here, $\mu$ and $\zeta$ are material-dependent lamé constants. See more details in Supplementary Material (SM)~\cite{supplemental}. The interlayer binding energy depends on both the interlayer spacing and the stacking configuration~\cite{lebedeva_interlayer_2011,verhoeven_model_2004},
\begin{equation}\label{eq:LB}
        L_{B}=V_{1}(d_{z}^{-}) + \sum_{j \in \{1,3,5\}} 2V_{B}V_{0}(d_{z}^{-})
        \cos(\pmb{g}_{j}\cdot\pmb{r}+\pmb{G}_{j}\cdot \pmb{u}^{-}).
\end{equation}
Here, $d_{z}^{-}=d_{0}+u_{z}^{-}$ is the local interlayer spacing with $d_{0}$ the interlayer spacing  in the rigid case. $V_{1}(d_{z}^{-})$ and $V_{0}(d_{z}^{-})$ depend solely on $d_z^{-}$. $V_{1}(d_{z}^{-})$ incorporates both van der Waals attraction and repulsion, and $V_{0}(d_{z}^{-})$ decays with increasing $d_z^{-}$ because the influence of stacking configuration on binding energy diminishes rapidly (see details in SM~\cite{supplemental}). $\pmb{G}_{j}$ is the monolayer reciprocal lattice vector. We define $\pmb{G}_{0}=0$, $\pmb{G}_{1}=\frac{4\pi}{\sqrt{3}a_{0}}\hat{\pmb{e}}_{y}$, and number the vectors counterclockwise, and $\pmb{g}_{j}$ are the moiré reciprocal vectors corresponding to $\pmb{G}_{j}$, here $a_0$ is the monolayer lattice constant.  Substituting the total Lagrangian density $L=L_{E}+L_{B}$ into Eq.~(\ref{eq:E-L eqs}) yields $\pmb{u}^{+}(\pmb{r})= 0$, meaning $\pmb{u}^{(t)}$ and $\pmb{u}^{(b)}$ are equal in magnitude but opposite in direction in the 2D whole space~\cite{supplemental}. We therefore focus on $\pmb{u}^{-}$. After detailed derivation, we obtain the following equations~\cite{supplemental},
\begin{equation}
    \begin{aligned}
        \Delta \pmb{u}^{-}_{\parallel} &= -\sum_{j\in \{1,3,5\}}
        \kappa_{\parallel}
        \sin(\pmb{g}_{j}\cdot\pmb{r})\pmb{G}_{j},\\
        \Delta \pmb{u}^{-}_{\perp} &= -\sum_{j\in \{1,3,5\}}
        \kappa_{\perp}
        \cos(\pmb{g}_{j}\cdot\pmb{r}) \hat{\pmb{e}}_{z},
    \end{aligned}
\end{equation}
where $\Delta=\frac{\partial^{2}}{\partial x^{2}}+\frac{\partial^{2}}{\partial y^{2}}+\frac{\partial^{2}}{\partial z^{2}}$ is the Laplacian. Due to the absence of explicit $z$-dependence for 2D systems, $\frac{\partial^{2}\pmb{u}^{-}_{\parallel}}{\partial z^{2}}=\frac{\partial^{2}\pmb{u}^{-}_{\perp}}{\partial z^{2}}=0$. Here, $\kappa_{\parallel}$ and $\kappa_{\perp}$ are material-dependent coefficients~\cite{supplemental}. Note that we treat $u_{z}^{-}$ as small compared to $d_{0}$ and neglect $\pmb{G}_{j}\cdot \pmb{u}^{-}_{\parallel}$, which has negligible impact except at extremely small angles (see SM~\cite{supplemental}).

\begin{figure}[t]
    \centering
    \includegraphics[width=0.5\textwidth]{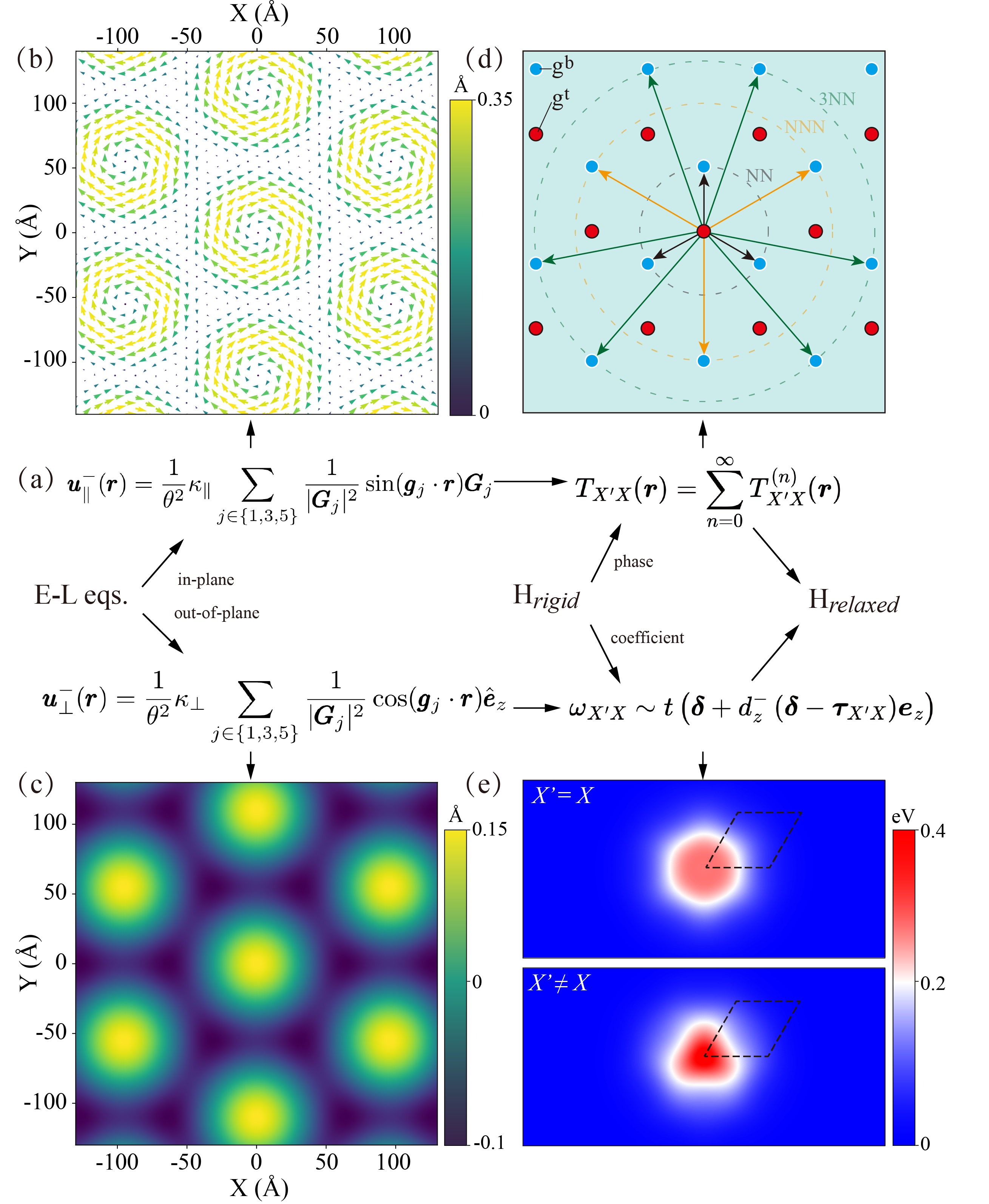}
    \caption{\label{fig:solveu} Relaxation and its effect on electronic structure. (a) Schematic workflow: Starting from Euler-Lagrange equations (E-L eqs.), we obtain relaxation configurations and analytical Hamiltonians for twisted systems. Real-space configuration of in-plane relaxation (b) and out-of-plane relaxation (c) in tMoTe$_{2}$ at 1.8$^{\circ}$. (d) The impact of in-plane relaxation on electronic structures manifests in phase factors, which we expand as hoppings in momentum space. Schematic for interlayer coupling Eq.~(\ref{eq:Tn_1}) at n=1 order. $\pmb{g}^{t}$ and $\pmb{g}^{b}$ denote the reciprocal lattice points of the top layer and the bottom layer, respectively. (e) Effect of out-of-plane relaxation on electronic structures via modified coupling coefficients, exemplified in TBG by the induced sublattice-dependent coefficient difference. The black dashed line indicates the unit cell of monolayer graphene.}
\end{figure}

Using the spectral method, we readily obtain the analytic solution
\begin{equation}\label{eq:solution}
    \begin{aligned}
        \pmb{u}_{\parallel}^{-}(\pmb{r}) &= \frac{1}{\theta^{2}}\kappa_{\parallel}\sum_{j\in \{1,3,5\}}\frac{1}{|\pmb{G}_{j}|^{2}} \sin(\pmb{g}_{j}\cdot\pmb{r})\pmb{G}_{j},\\
        \pmb{u}_{\perp}^{-}(\pmb{r}) &= \frac{1}{\theta^{2}}\kappa_{\perp} \sum_{j\in \{1,3,5\}}\frac{1}{|\pmb{G}_{j}|^{2}}
        \cos(\pmb{g}_{j}\cdot \pmb{r}) \hat{\pmb{e}}_{z}.
    \end{aligned}
\end{equation}
These analytical expressions are our intermediate result, with the in-plane relaxation component being consistent with results obtained from distinct approaches~\cite{ezzi_analytical_2024,ceferino_pseudomagnetic_2024}. For tMoTe${}_2$ at $1.8^{\circ}$, the real-space relaxation distribution is shown in Figs.~\ref{fig:solveu}(b)(c), which reproduces the key features of prior DFT, molecular dynamics and experimental results~\cite{uchida_atomic_2014,shabani_deep_2021,liuMoirePhononsMagicAngle2022,liuDPmoireToolConstructing2024,cantele_structural_2020}. The relaxation amplitude scales inversely with the square of the twist angle $\theta$. Equations~(\ref{eq:solution}) diverge as $\theta \to 0$, hence the solution is valid only above a critical angle $\theta^{*}$ (typically $\theta^{*} < 1^{\circ}$; see SM~\cite{supplemental}). We propose a simple yet effective remedy: replacing $1/\theta^{2}$ with $1/\max\{\theta^{2},\theta^{*2}\}$ in Eq.~(\ref{eq:solution}). This avoids divergence while preserving the correct relaxation behavior.

\textit{Numerical Solution for Relaxation.---}
The analytic solution fails at extremely small twist angles because we neglected the $\pmb{G}_{j}\cdot \pmb{u}^{-}_{\parallel}$ term in Eq.~(\ref{eq:LB}), which exerts a certain influence under such a small-angle limit. The complete form of the equations is
\begin{equation}\label{eq:num_eq}
    \begin{aligned}
        \Delta \pmb{u}^{-}_{\parallel} &= -\sum_{j\in \{1,3,5\}}
        \kappa_{\parallel}
        \sin(\pmb{g}_{j}\cdot\pmb{r}+\pmb{G}_{j}\cdot \pmb{u}^{-}_{\parallel})\pmb{G}_{j},\\
        \Delta \pmb{u}^{-}_{\perp} &= -\sum_{j\in \{1,3,5\}}
        \kappa_{\perp}
        \cos(\pmb{g}_{j}\cdot\pmb{r}+\pmb{G}_{j}\cdot \pmb{u}^{-}_{\parallel}) \hat{\pmb{e}}_{z}.
    \end{aligned}
\end{equation}
These equations admit no analytic solution. Here, we propose an iterative numerical approach to solve them. Considering the in-plane relaxation, we replace $\pmb{G}_{j}\cdot \pmb{u}^{-}_{\parallel}$ with $\frac{n}{N} \pmb{G}_{j}\cdot \pmb{u}^{-}_{\parallel}$, where $n\in \{0,1,\cdots, N\}$. The analytic solution for $n=0$ is already known. Substituting this into the right-hand side of the equation for $n=1$ yields a new solution. Iterating this process, we obtain
\begin{equation}
    \Delta \pmb{u}^{-,(n)}_{\parallel} = -\sum_{j\in \{1,3,5\}}
        \kappa_{\parallel}
        \sin\left(\pmb{g}_{j}\cdot\pmb{r}+ \frac{n}{N} \pmb{G}_{j}\cdot \pmb{u}^{-,(n-1)}_{\parallel}\right)\pmb{G}_{j}.
\end{equation}
For suﬀiciently large N, using the spectral method, we can iteratively solve the above equation to obtain the accurate numerical solution of the in-plane relaxation (see SM for details~\cite{supplemental}). Substituting this solution into Eq.~(\ref{eq:num_eq}), we then derive the numerical solution of the out-of-plane relaxation.

\textit{Impact of Relaxation on Electronic Structure.---}
Considering the $\pmb{K}_{\eta}$ valley in a hexagonal Bravais lattice, we can write the Hamiltonian for an untwisted bilayer system as follows~\cite{lopesdossantos_graphene_2007a,bistritzer_moire_2011,jung_initio_2014a,wu_topological_2019}
\begin{equation}\label{eq:untwisted}
    H_{\eta}(\pmb{k}, \pmb{\delta})= 
    \left(
    \begin{matrix}
        h_{\eta}(\pmb{k})+V^{t}(\pmb{\delta}) &  T(\pmb{\delta}) \\
        T^{\dagger}(\pmb{\delta}) & h_{\eta}(\pmb{k})+V^{b}(\pmb{\delta})
    \end{matrix}
    \right).
\end{equation}
Here, $\eta$ is the valley index, $\pmb{\delta}$ is the in-plane interlayer displacement, $h_{\eta}(\pmb{k})$ is the monolayer Hamiltonian, $V^{l}(\pmb{\delta})$ represents intralayer potential that varies with $\pmb{\delta}$, and $T(\pmb{\delta})$ denotes interlayer coupling. Their forms are
\begin{equation}
    \begin{aligned}
        V^{l}_{X'X}(\pmb{\delta})&=\sum_{i=1}^{6}\nu_{X'X}^{l,i}e^{i\pmb{G}_{i}\cdot \pmb{\delta}}, \\
        T_{X'X}(\pmb{\delta})
        &=\sum_{i\in \{0,2,3\}}\omega_{X'X}
        e^{i\pmb{G}_i\cdot\pmb{\tau}_{X'X}}e^{i(\pmb{G}_i+\pmb{K}_{\eta})\cdot\pmb{\delta}},
    \end{aligned}
\end{equation}
where $X,X'$ are orbital indices, and $\pmb{\tau}_{X'X}$ is the relative position vector between the two orbitals. The coupling coefficients can be expressed as integrals~\cite{supplemental}
\begin{equation}
    \begin{aligned}
        \nu_{X'X}^{l,i}=&\frac{1}{S_0}\sum_{j}e^{-i\pmb{K}_{\eta}\cdot(\pmb{R}_{j}+\pmb{\tau}_{X'X})}   \\
        &\times \int_{S_{0}}h^{l}_{X'X}(\pmb{R}_{j};\pmb{\delta}, d_{z}^{-}(\pmb{\delta}))e^{-i\pmb{G}_{i}\cdot \pmb{\delta}}d\pmb{\delta},\\
        \omega_{X'X}=&-\frac{1}{S_0}\int_{\infty}
    t(\pmb{\delta}+d_{z}^{-}(\pmb{\delta}-\pmb{\tau}_{X'X})\pmb{e}_{z})
    e^{-i\pmb{K}_{\eta}\cdot\pmb{\delta}}
    d\pmb{\delta}.
    \end{aligned}
\end{equation}
For the intralayer part, $h^{l}_{X'X}(\pmb{R}_{i};\pmb{\delta}, d_{z}^{-}(\pmb{\delta}))$ represents the hopping between orbitals $X'$ and $X$ separated by $\pmb{R}_{i}$ in layer $l$~\cite{wu_topological_2019, supplemental}. The functional form can be obtained from DFT calculations of untwisted bilayer structures combined with Wannier functions and is a functional of $d_{z}^{-}(\pmb{\delta})=u_{z}^{-}(\pmb{\delta})+d_{0}$. For the interlayer part, given the weak van der Waals interactions, we adopt the two-center approximation, where $t(\pmb{\delta}+d_{z}^{-}(\pmb{\delta}-\pmb{\tau}_{X'X})\pmb{e}_{z})$ takes the Slater-Koster form~\cite{moon_optical_2013,koshino_maximally_2018}, also a functional of $d_{z}^{-}(\pmb{\delta})$. In TBG, out-of-plane relaxation causes 
$t(\pmb{\delta}+d_{z}^{-}(\pmb{\delta}-\pmb{\tau}_{X'X})\pmb{e}_{z})$ to adopt distinct functiona forms when 
$X^{'}$ and $X$ denote identical orbitals versus different orbitals, as shown in Fig. \ref{fig:solveu}(e), thereby inducing critical differences in coupling coefficients. As described above, we incorporate out-of-plane relaxation in the coefficients integrals; its influence mainly manifests in the coupling coefficients.

Next, we address in-plane relaxation. The core concept in constructing continuum models for twisted systems is the local stacking approximation. In the rigid case, the local interlayer displacement arises from twisting: $\pmb{\delta}=\pmb{d}(\theta)=[R(\theta/2)-R(-\theta/2)]\pmb{r}$ in Eq (\ref{eq:untwisted}). Thus, $\pmb{G}_{i}\cdot \pmb{\delta}=\pmb{g}_{i}\cdot \pmb{r}$ and $\pmb{K}_{\eta}\cdot \pmb{\delta}=\pmb{q}_{\eta}\cdot \pmb{r}$. However, with in-plane relaxation, local displacement includes both twist-induced and relaxation-induced components: $\pmb{\delta}=\pmb{d}(\theta)+\pmb{u}^{-}_{\parallel}$. Consequently, $\pmb{G}_{i}\cdot \pmb{\delta}=\pmb{g}_{i}\cdot \pmb{r}+\pmb{G}_{i}\cdot \pmb{u}^{-}_{\parallel}$ and $\pmb{K}_{\eta}\cdot \pmb{\delta}=\pmb{q}_{\eta}\cdot \pmb{r}+\pmb{K}_{\eta}\cdot \pmb{u}^{-}_{\parallel}$. The coupling terms then become
\begin{equation}\label{eq:VT_Gu}
    \begin{aligned}
        V^{l}_{X'X}(\pmb{r})&=\sum_{i=1}^{6}\nu_{X'X}^{l,i}e^{i\pmb{g}_{i}\cdot \pmb{r}}e^{i\pmb{G}_{i}\cdot \pmb{u}^{-}_{\parallel}}, \\
        T_{X'X}(\pmb{r})
        &=\sum_{i\in \{0,2,3\}}\omega_{X'X}
        e^{i\pmb{G}_i\cdot\pmb{\tau}_{X'X}}e^{i(\pmb{g}_i+\pmb{q}_{\eta})\cdot\pmb{r}}e^{i(\pmb{G}_{i}+\pmb{K}_{\eta})\cdot \pmb{u}^{-}_{\parallel}}.
    \end{aligned}
\end{equation}
For convenience, we define $\pmb{Q}_{i}$, for intralayer moiré potential, $\pmb{Q}_{i}:=\pmb{G}_{i}$; for interlayer coupling, $\pmb{Q}_{i}:=\pmb{G}_{i}+\pmb{K}_{\eta}$. It can be seen that, compared with the Hamiltonian in the rigid case, in-plane relaxation introduces a phase factor $e^{i\pmb{Q}_{i}\cdot \pmb{u}^{-}_{\parallel}}$, which prevents direct solution of the Hamiltonian.

With our analytic solution for in-plane relaxation, we directly evaluate the exponent
\begin{equation}
    \begin{aligned}
        i\pmb{Q}_i\cdot \pmb{u}^{-}_{\parallel}&=i\pmb{Q}_{i} \cdot \frac{1}{\theta^{2}}\kappa_{\parallel}\sum_{j\in \{1,3,5\}}\frac{1}{|\pmb{G}_{j}|^{2}} \sin(\pmb{g}_{j}\cdot\pmb{r})\pmb{G}_{j}\\
        &=\sum_{j=1}^{6}\frac{\kappa_{\parallel}}{2}\frac{1}{\theta^{2}}\frac{\pmb{Q}_{i}}{|\pmb{G}_{j}|} \cdot \hat{\pmb{G}}_{j}e^{i\pmb{g}_j \cdot \pmb{r}}=\sum_{j=1}^{6}\gamma_{ij}e^{i\pmb{g}_j \cdot \pmb{r}},
    \end{aligned}
\end{equation}
with definition $\gamma_{ij}:=\frac{\kappa_{\parallel}}{2}\frac{1}{\theta^{2}}\frac{(\pmb{G}_{i}+\pmb{K}_{\eta})}{|\pmb{G}_{j}|} \cdot \hat{\pmb{G}}_{j}$ for interlayer coupling and $\gamma_{ij}:=\frac{\kappa_{\parallel}}{2}\frac{1}{\theta^{2}}\frac{\pmb{G}_{i}}{|\pmb{G}_{j}|} \cdot \hat{\pmb{G}}_{j}$ for intralayer moiré potential. The factor can be expanded into series as follows
\begin{equation}
    \begin{aligned}
        e^{i\pmb{Q}_{i}\cdot \pmb{u}^{-}_{\parallel}} =& \sum_{n=0}^{\infty}\frac{(i\pmb{Q}_{i}\cdot \pmb{u}^{-}_{\parallel})^{n}}{n!}\\
        =&\sum_{n=0}^{\infty}\sum_{j_1j_2\cdots j_n=1}^{6}\frac{\prod_{m=1}^{n}\gamma_{ij_{m}}}{n!}
        e^{i(\sum_{m=1}^{n}\pmb{g}_{j_{m}}) \cdot \pmb{r}} \\
        =&1 +  \sum_{j=1}^{6}\gamma_{ij}e^{i\pmb{g}_{j}\cdot \pmb{r}}
        +\sum_{j_{1},j_{2}=1}^{6}\frac{\gamma_{ij_{1}}\gamma_{ij_{2}}}{2}e^{i(\pmb{g}_{j_{1}}+\pmb{g}_{{j_{2}}})\cdot \pmb{r}}\\ &+\cdots .
    \end{aligned}
\end{equation}
\begin{figure}[t]
    \centering
    \includegraphics[width=0.49\textwidth]{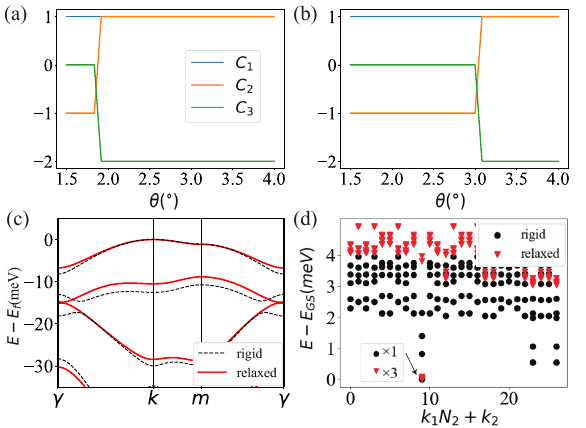}
    \caption{\label{fig:topo} 
    The effect of relaxation on the electronic structure of tMoTe$_{2}$. (a)-(b) Evolution of topology for the first three bands in tMoTe$_{2}$ with twist angle $\theta$. (a) $n=0$: pure rigid case, phase transition occurs at $\sim 1.8^{\circ}$. (b) $n=2$: including first and second order in-plane relaxation effects, phase transition shifts to $\sim 3^{\circ}$. (c) The  band structures of tMoTe$_{2}$, with $\theta=3^{\circ}$. (d) Many-body energy spectra for rigid and relaxed case at hole filling $\nu=\frac{2}{3}$, with $\theta=3^{\circ}$.
    }
\end{figure}
Thus, combining Eq.~(\ref{eq:VT_Gu}), the moiré potential and interlayer coupling terms become
\begin{equation}
    \begin{aligned}
        V_{X'X}^{l}(\pmb{r})
        &=\sum_{n=0}^{\infty}V_{X'X}^{l,(n)}(\pmb{r}),\\
        T_{X'X}(\pmb{r})
        &=\sum_{n=0}^{\infty}T_{X'X}^{(n)}(\pmb{r}),
    \end{aligned}
\end{equation}
where $n$ denotes the $n$-th order effect of in-plane relaxation. Specifically,
\begin{equation}
    \begin{aligned}
        V_{X'X}^{l,(n)}(\pmb{r})=&\sum_{i=1}^{6}\sum_{j_1j_2\cdots j_n=1}^{6}
        \nu_{X'X}^{l,i} \times \\
        &\frac{\prod_{m=1}^{n}\gamma_{ij_{m}}}{n!}
        \exp\left[i\left(\pmb{g}_i+\sum_{m=1}^n\pmb{g}_{j_m}\right)\cdot\pmb{r}\right],
    \end{aligned}
\end{equation}
\begin{equation}\label{eq:inter_inplane}
    \begin{aligned}
        T_{X'X}^{(n)}(\pmb{r})=&\sum_{i\in \{0,2,3\}}\sum_{j_1j_2\cdots j_n=1}^{6}
        \omega_{X'X}e^{i\pmb{G}_i\cdot\pmb{\tau}_{X'X}} \times \\
        &\frac{\prod_{m=1}^{n}\gamma_{ij_{m}}}{n!}
        \exp\left[i\left(\pmb{q}_{\eta}+\pmb{g}_i+\sum_{m=1}^{n}\pmb{g}_{j_m}\right)\cdot\pmb{r}\right].
    \end{aligned}
\end{equation}
These expressions represent the main results of our work. We transform the in-plane relaxation in the phase factor into couplings between different reciprocal lattice sites. The $n=0$ term corresponds to the rigid case. For $n=1$, the moiré potential and interlayer coupling are
\begin{equation}
    V_{X'X}^{l,(1)}(\pmb{r}) = \sum_{i=1}^{6} \sum_{j=1}^{6} \nu_{X'X}^{l,i} \gamma_{ij} e^{i (\pmb{g}_i + \pmb{g}_j) \cdot \pmb{r}},
\end{equation}
\begin{equation}\label{eq:Tn_1}
    T^{(1)}_{X'X}(\pmb{r}) = \sum_{i \in \{0,2,3\}} \sum_{j=1}^{6} \omega_{X'X} \gamma_{ij} e^{i \pmb{G}_i \cdot \pmb{\tau}_{X'X}} e^{i (\pmb{q}_{\eta} + \pmb{g}_i + \pmb{g}_j) \cdot \pmb{r}}.
\end{equation}
Taking the interlayer coupling as an example, considering the case of n=1 (i.e., Eq.~(\ref{eq:Tn_1})), we schematically illustrate it in Fig.~\ref{fig:solveu}(d). $T^{(1)}_{X'X}(\pmb{r})$ contains 18 hopping terms in reciprocal space, where both the nearest-neighbor (NN) and next-nearest-neighbor (NNN) hoppings are counted twice. Their amplitude is $\omega_{X'X}\gamma_{ij}$, and due to the presence of $\gamma_{ij}$, this amplitude depends on both the direction and distance of the hopping and the twist angle $\theta$. Higher-order terms follow analogously.

The series expansion approach for the phase factor necessitates examining convergence. For convergence of the in-plane relaxation coupling terms, we require $\frac{\gamma_{ij}}{\sqrt[n]{n!}}<1$. Conservatively, we impose $\max(\gamma_{ij})<1$, i.e., $\frac{\kappa_{\parallel}}{2}\frac{1}{\theta^{2}}<1$, yielding the twist angle constraint
\begin{equation}
    \theta > \sqrt{\frac{\kappa_{\parallel}}{2}}. 
\end{equation}
Thus, our theory is applicable above a critical angle $\theta^{\dagger}=\sqrt{\frac{\kappa_{\parallel}}{2}}$. Typically, $\theta^{\dagger}\approx 0.3^{\circ}$ for graphene and $\theta^{\dagger}\approx 1^{\circ}$ for TMDs~\cite{supplemental}.

As an application of our theoretical framework, we computed the topological properties of tMoTe$_{2}$. As shown in Figs.~\ref{fig:topo}(a)(b), in the $n=0$ case (fully rigid limit), the system undergoes a topological phase transition with changing Chern numbers from $(1,-1,0)$ to $(1,1,-2)$ at $1.8^{\circ}$. When full relaxation is considered analytically, the transition shifts to $3^{\circ}$ (convergence is achieved at $n=2$), consistent with DFT calculations~\cite{mao_transfer_2024a,jia_moire_2024a,yu_fractional_2024,wang_fractional_2024,xu_maximally_2024}. The band structures are shown in Fig.~\ref{fig:topo}(c). Parameters for the case without in-plane relaxation are adopted from Ref.~\cite{wu_topological_2019}. 
Relaxation also exerts a non-negligible influence on the many-body electronic properties of the system. We compared the many-body energy spectra of rigid and relaxed tMoTe$_2$ at hole filling $\nu = \frac{2}{3}$ using exact diagonalization of the 27-site cluster case\cite{supplemental}, as shown in Fig.~\ref{fig:topo}(d). In the rigid case, the system has no many-body energy gap, whereas under relaxation, there are three degenerate states concentrated at the Brillouin zone center. Combined with many-body Chern number calculations, we conclude that at the twist angle of $\theta = 3^\circ$, the system becomes a FCI when relaxation is taken into account~\cite{supplemental}. The characteristic angle at which the FCI emerges is also close to experimental observations~\cite{zengThermodynamicEvidenceFractional2023}.

\begin{figure}[t]
    \centering
    \includegraphics[width=0.48\textwidth]{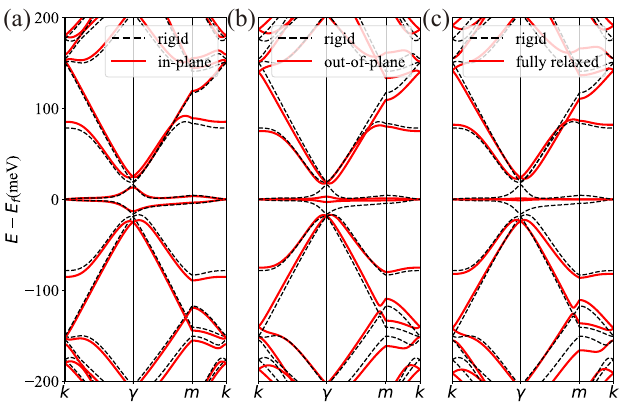}
    \caption{\label{fig:TBG} Band structure of 1.05° TBG under relaxation: (a) considering only in-plane relaxation, (b) considering only out-of-plane relaxation, (c) considering both relaxation completely.}
\end{figure}

Applying the framework to TBG, Fig.~\ref{fig:TBG} demonstrates that near the magic angle ($\theta=1.05^{\circ}$): (i) Rigid bands exhibit degeneracy between low-energy and excited states; (ii) Introducing either relaxation mechanism lifts this degeneracy and reduces bandwidth; (iii) Full relaxation yields nearly exactly flat bands.
This can be explained by stacking-region contribution imbalance: in-plane relaxation shrinks the AA region while expanding the AB region; Out-of-plane relaxation increases AA interlayer spacing (weakening coupling) while decreasing AB spacing (enhancing coupling). Both effects weaken the contribution from the AA regions while enhancing that from the AB regions, driving the system closer to the chiral limit (i.e., the exactly flat band limit), which is also consistent with DFT calculations and experimental results~\cite{lisi_observation_2021,cantele_structural_2020,lucignano_crucial_2019,kazmierczak_strain_2021,tarnopolsky_origin_2019}. 

\textit{Conclusion.---}
This work establishes a systematic analytical framework for lattice relaxation in twisted systems. We derive a closed-form solution for relaxation displacement fields from continuum elastic theory and employ a phase factor expansion theory to treat the effects of in-plane relaxation on the electronic structures. Combined with analytical treatment of out-of-plane relaxation, this provides a complete description of relaxation effects in twisted systems.

Applying this framework to tMoTe$_{2}$, we successfully reproduce relaxation-mediated topological phase transitions. Our predicted transition point aligns DFT results, resolving the discrepancy from rigid models. Furthermore, many-body calculations at fractional filling $\nu = \frac{2}{3}$ confirm that relaxation stabilizes the fractional Chern insulator state, consistent with experimental observations. Applying the framework to TBG, we clearly reproduce relaxation's modulation of flat bands—lifting low-energy degeneracies to form isolated flat bands, providing intuitive insights into the origin of flat band.

The framework's core value lies in: Transcending the limitations of traditional DFT by establishing the analytical physical picture of relaxation-electron coupling. Computational efficiency improves by over four orders of magnitude, enabling comprehensive parameter space exploration. It is worth noting that this work focuses primarily on hexagonal Bravais lattice systems, which are the most studied in twistronics. As a general method, our theory is equally applicable to other systems, requiring only modifications to the forms of energy function Eq.~(\ref{eq:LE}) and Eq.~(\ref{eq:LB}).

Current limitations include: Failure of analytic solutions and series expansions at extremely small angles ($\theta < \theta^{*} \text{ and } \theta^{\dagger}$) due to emergent strong relaxation self-coupling effects, necessitating the development of non-perturbative theories for extremely small twist angles. Moreover, the discrepancies among previously reported DFT results regarding the phase transition point in tMoTe$_{2}$ suggest that possible additional factors need to be considered.

This work provides an analytical perspective on relaxation-electron coupling in twisted systems, advancing research from numerical fitting to an analytical paradigm. It establishes a theoretical foundation for emerging novel physics and designing quantum functional devices based on moiré superlattices.

\begin{acknowledgments}
\textit{Acknowledgments.}--- The authors thank Jingyi Duan for helpful discussions. The work is supported by the NSF of China (Grant No. 12374055), and the Science Fund for Creative Research Groups of NSFC (Grant No. 12321004), the National Key R\&D Program of China (Grant No. 2020YFA0308800).
\end{acknowledgments}

\bibliography{ref}

\end{document}


\title{Supplementary Material for ``Relaxation and Its Effects on Electronic Structure in Twisted Systems: An Analytical Perspective"}
\author{Junxi Yu}
\author{Bingbing Wang}
\author{Cheng-Cheng Liu}
\email{ccliu@bit.edu.cn}
\affiliation{Centre for Quantum Physics, Key Laboratory of Advanced Optoelectronic Quantum Architecture and Measurement (MOE), School of Physics, Beijing Institute of Technology, Beijing 100081, China}
\maketitle
\tableofcontents
\newpage

\section{ Derivation of the Relaxation Equations}
In this section, we derive the relaxation equations for twisted bilayer systems starting from the Euler-Lagrange equations and the system's total energy.
\subsection{ The Euler-Lagrange Equations}
We define the local displacement field due to relaxation in the $l$th layer of the twisted bilayer as
\begin{equation}
    \pmb{u}^{(l)}(\pmb{r})=\pmb{u}^{(l)}(x,y)=u^{(l)}_{x}(x,y)\hat{\pmb{e}}_{x}+u^{(l)}_{y}(x,y)\hat{\pmb{e}}_{y}+u^{(l)}_{z}(x,y)\hat{\pmb{e}}_{z}.
\end{equation}
It is a function of position $\pmb{r}=x\hat{\pmb{e}}_{x}+y\hat{\pmb{e}}_{y}$. Since the systems we consider are two-dimensional (2D), $\pmb{u}^{(l)}(\pmb{r})$ does not explicitly depend on $z$. Here, $\hat{\pmb{e}}_{i}$ denotes the unit vector along the $i$-axis ($i \in \{x, y, z\}$), and $l=t,b$ label the top layer and bottom layer.

For a bilayer system, we define the transformation
\begin{equation}\label{eq:transform}
    \begin{aligned}
        \pmb{u}^{+}(\pmb{r})&=\pmb{u}^{(t)}(\pmb{r})+\pmb{u}^{(b)}(\pmb{r}),\\
        \pmb{u}^{-}(\pmb{r})&=\pmb{u}^{(t)}(\pmb{r})-\pmb{u}^{(b)}(\pmb{r}),
    \end{aligned}
\end{equation}
with the inverse transformation
\begin{equation}\label{eq:inverse_transform}
    \begin{aligned}
        \pmb{u}^{(t)}&=\frac{\pmb{u}^{+}+\pmb{u}^{-}}{2},\\
        \pmb{u}^{(b)}&=\frac{\pmb{u}^{+}-\pmb{u}^{-}}{2}.
    \end{aligned}
\end{equation}
The transformed displacement fields also have the form
\begin{equation}
    \pmb{u}^{\pm}(\pmb{r})=\pmb{u}^{\pm}(x,y)=u^{\pm}_{x}(x,y)\hat{\pmb{e}}_{x}+u^{\pm}_{y}(x,y)\hat{\pmb{e}}_{y}+u^{\pm}_{z}(x,y)\hat{\pmb{e}}_{z}.
\end{equation}
We can decompose them into in-plane and out-of-plane components
\begin{equation}\label{eq:upm}
    \pmb{u}^{\pm}(\pmb{r})=\pmb{u}^{\pm}_{\parallel}(\pmb{r})+\pmb{u}^{\pm}_{\perp}(\pmb{r}),
\end{equation}
where $\pmb{u}^{\pm}_{\parallel}(\pmb{r})=u^{\pm}_{x}(x,y)\hat{\pmb{e}}_{x}+u^{\pm}_{y}(x,y)\hat{\pmb{e}}_{y}$ and $\pmb{u}^{\pm}_{\perp}(\pmb{r})=u^{\pm}_{z}(x,y)\hat{\pmb{e}}_{z}$.

The Lagrangian density depends on the fields $\pmb{u}^{\pm}(\pmb{r})$ as
\begin{equation}
    \begin{aligned}
        L(&x,y,z;u_{x}^{+},u_{y}^{+},u_{z}^{+};u_{x}^{-},u_{y}^{-},u_{z}^{-};\\
        &\partial_{x}u_{x}^{+},\partial_{x}u_{y}^{+},\partial_{x}u_{z}^{+};\partial_{x}u_{x}^{-},\partial_{x}u_{y}^{-},\partial_{x}u_{z}^{-};\\
        &\partial_{y}u_{x}^{+},\partial_{y}u_{y}^{+},\partial_{y}u_{z}^{+};\partial_{y}u_{x}^{-},\partial_{y}u_{y}^{-},\partial_{y}u_{z}^{-};\\
        &\partial_{z}u_{x}^{+},\partial_{z}u_{y}^{+},\partial_{z}u_{z}^{+};\partial_{z}u_{x}^{-},\partial_{z}u_{y}^{-},\partial_{z}u_{z}^{-}).
        \end{aligned}
\end{equation}
The total energy $U$ is given by the integral of the Lagrangian density
\begin{equation}
    U=\int L \, dx\,dy\,dz.
\end{equation}

It should be clarified here that the Lagrangian density $L$ is generally defined as the difference between the kinetic energy density and the potential energy density. However, this work deals with static relaxation problems, where the kinetic energy term is zero. Therefore, the Lagrangian density in this work is actually equal to the negative of the potential energy density, i.e., $L = -V$. Consequently, the quantity $U = \int L d\pmb{r}$ is equal to the negative of the total potential energy ($-E_{\text{pot}}$). Since the Euler-Lagrange equation is derived from the action variation principle $\delta S = 0$, and constant scalar factors (such as the negative sign) do not affect the position of the extreme point, the equilibrium equation obtained by variational derivation using $U$ is completely equivalent to the variation directly performed on the total potential energy $E_{\text{pot}}$.

Varying the total energy yields
\begin{equation}
    \delta U = \int
    \sum_{i\in\{x,y,z\}}\sum_{\pm}\left[\frac{\partial L}{\partial u_{i}^{\pm}}\delta u_{i}^{\pm}+
    \sum_{j\in\{x,y,z\}}
    \frac{\partial L}{\partial (\partial_{j} u_{i}^{\pm})}\delta (\partial_{j} u_{i}^{\pm})\right]
    \,dx\,dy\,dz.
\end{equation}
Since variation and differentiation commute ($\delta (\partial_{j} u_{i}^{\pm}) = \partial_{j} (\delta u_{i}^{\pm})$), integrating by parts and setting boundary terms to zero gives
\begin{equation}
    \delta U = \int
    \sum_{i\in\{x,y,z\}}\sum_{\pm}\left[\frac{\partial L}{\partial u_{i}^{\pm}}-
    \sum_{j\in\{x,y,z\}}
    \frac{\partial}{\partial r_{j}}
    \frac{\partial L}{\partial (\partial_{j} u_{i}^{\pm})}\right]\delta u_{i}^{\pm}
    \,dx\,dy\,dz.
\end{equation}
Requiring $\delta U = 0$ for arbitrary variations $\delta u_{i}^{\pm}$ leads to two sets of Euler-Lagrange equations. For $\pmb{u}^{+}$
\begin{equation}\label{eq:Lup}
    \begin{aligned}
        &\frac{\partial L}{\partial u_{x}^{+}}- \frac{\partial}{\partial x}\frac{\partial L}{\partial (\partial_{x} u_{x}^{+})} - \frac{\partial}{\partial y} \frac{\partial L}{\partial (\partial_{y} u_{x}^{+})} - \frac{\partial}{\partial z}\frac{\partial L}{\partial (\partial_{z} u_{x}^{+})} =      0 \\
        &\frac{\partial L}{\partial u_{y}^{+}}- \frac{\partial}{\partial x}\frac{\partial L}{\partial (\partial_{x} u_{y}^{+})} - \frac{\partial}{\partial y} \frac{\partial L}{\partial (\partial_{y} u_{y}^{+})} - \frac{\partial}{\partial z}\frac{\partial L}{\partial (\partial_{z} u_{y}^{+})} = 0 \\
        &\frac{\partial L}{\partial u_{z}^{+}}- \frac{\partial}{\partial x}\frac{\partial L}{\partial (\partial_{x} u_{z}^{+})} - \frac{\partial}{\partial y} \frac {\partial L}{\partial (\partial_{y} u_{z}^{+})} - \frac{\partial}{\partial z}\frac{\partial L}{\partial (\partial_{z} u_{z}^{+})} = 0. \\
    \end{aligned}
\end{equation}
For $\pmb{u}^{-}$
\begin{equation}\label{eq:Lun}
    \begin{aligned}
        &\frac{\partial L}{\partial u_{x}^{-}}- \frac{\partial}{\partial x}\frac{\partial L}{\partial (\partial_{x} u_{x}^{-})} - \frac{\partial}{\partial y}\frac{\partial L}{\partial (\partial_{y} u_{x}^{-})} - \frac{\partial}{\partial z}\frac{\partial L}{\partial (\partial_{z} u_{x}^{-})} = 0 \\
        &\frac{\partial L}{\partial u_{y}^{-}}- \frac{\partial}{\partial x}\frac{\partial L}{\partial (\partial_{x} u_{y}^{-})} - \frac{\partial}{\partial y}\frac{\partial L}{\partial (\partial_{y} u_{y}^{-})} - \frac{\partial}{\partial z}\frac{\partial L}{\partial (\partial_{z} u_{y}^{-})} = 0 \\
        &\frac{\partial L}{\partial u_{z}^{-}}- \frac{\partial}{\partial x}\frac{\partial L}{\partial (\partial_{x} u_{z}^{-})} - \frac{\partial}{\partial y}\frac{\partial L}{\partial (\partial_{y} u_{z}^{-})} - \frac{\partial}{\partial z}\frac{\partial L}{\partial (\partial_{z} u_{z}^{-})} = 0. \\
    \end{aligned}
\end{equation}

\subsection{ Total Energy and Relaxation Equations}
The total energy $U$ in a twisted bilayer consists of three parts: the elastic energy of the top and bottom layers $U_{E}^{(l)}$, and the interlayer binding energy $U_{B}$,
\begin{equation}
    U = U_{B} + \sum_{l} U_{E}^{(l)}.
\end{equation}

First, consider the elastic energy. For a 2D hexagonal Bravais lattice, common in twisted systems, the elastic energy per layer based on crystal elasticity theory \cite{landau1986} is
\begin{equation}
    \begin{aligned}
        U_{E}^{(l)} &=
        \int
        \Bigg\{\frac{1}{2}\lambda_{1}(u^{(l)}_{zz})^{2} + 2\lambda_{2}(u^{(l)}_{xx} + u^{(l)}_{yy})^{2} + \lambda_{3}\left[(u^{(l)}_{xx} - u^{(l)}_{yy})^{2} + 4(u^{(l)}_{xy})^{2}\right] \\
        &+ 2\lambda_{4}u^{(l)}_{zz}(u^{(l)}_{xx} + u^{(l)}_{yy}) + 4\lambda_{5}\left((u^{(l)}_{xz})^{2} + (u^{(l)}_{yz})^{2}\right) \Bigg\} d \pmb{r},
        \end{aligned}
\end{equation}
here $\lambda_{1-5}$ are the lamé constants. It is noteworthy that, due to the emergent symmetry characteristics of moiré systems, the relaxation patterns and their effects are essentially identical for systems with $D_{3}$ and $D_{6}$ symmetries\cite{angeli_valley_2021}. Here,
\begin{equation}
    u_{ik}^{(l)} = \frac{1}{2}\left(\frac{\partial u_{i}^{(l)}}{\partial r_{k}} + \frac{\partial u_{k}^{(l)}}{\partial r_{i}}\right),
\end{equation}
where $i, k \in \{x, y, z\}$ and $l$ is the layer index. Note that $\pmb{u}^{(l)}(\pmb{r})$ is the displacement field, while $u_{ik}^{(l)}$ is the strain tensor; they represent distinct physical quantities.

Both experiments and molecular dynamics calculations demonstrate that local twisting plays the completely dominant role in relaxation, while local volume compression is negligible\cite{ezzi_analytical_2024,ceferino_pseudomagnetic_2024,kazmierczak_strain_2021}. Therefore, the zero trace condition is employed
\begin{equation}
    \nabla \cdot \pmb{u}^{(l)}(\pmb{r}) = 0.
\end{equation}
This simplifies the elastic energy to
\begin{equation}
    U_{E}^{(l)} =
\int\left\{\lambda_{3}\left[(u^{(l)}_{xx} - u^{(l)}_{yy})^{2} + 4(u^{(l)}_{xy})^{2}\right] + 4\lambda_{5}\left[(u^{(l)}_{xz})^{2} + (u^{(l)}_{yz})^{2}\right] \right\} d \pmb{r}.
\end{equation}
Notably, for corrugated 2D materials where the displacement field does not explicitly depend on $z$, we have
\begin{equation}
    u^{(l)}_{xz} = \frac{1}{2}\frac{\partial u^{(l)}_{z}}{\partial x}, \quad u^{(l)}_{yz} = \frac{1}{2}\frac{\partial u^{(l)}_{z}}{\partial y}.
\end{equation}
For convenience, we redefine $\lambda_3 = \frac{\mu}{2}$ and $\lambda_5 = \zeta$. Applying the transformation Eq.~(\ref{eq:transform}), $\sum_{l}U_{E}^{(l)} = \sum_{\pm}U_{E}^{\pm}$, where
\begin{equation}
    U_E^{\pm} =
\int
\left\{
\frac{\mu}{4}\left[(u_{xx}^{\pm} - u_{yy}^{\pm})^{2} + 4(u_{xy}^{\pm})^{2}\right]
+ 2\zeta\left[(u_{xz}^{\pm})^{2} + (u_{yz}^{\pm})^{2}\right]
\right\}
d \pmb{r},
\end{equation}
which expands to
\begin{equation}
    U_E^{\pm}
= \int
\left\{
\frac{\mu}{4}\left[\left(\frac{\partial u_{x}^{\pm}}{\partial x} - \frac{\partial u_{y}^{\pm}}{\partial y}\right)^{2} + \left(\frac{\partial u_{x}^{\pm}}{\partial y} + \frac{\partial u_{y}^{\pm}}{\partial x}\right)^{2}\right]
+
\frac{\zeta}{2}\left[\left(\frac{\partial u_{z}^{\pm}}{\partial x}\right)^{2} + \left(\frac{\partial u_{z}^{\pm}}{\partial y}\right)^{2}\right]
\right\}
d \pmb{r}.
\end{equation}

\begin{figure}[t]
    \centering
    \includegraphics[width=0.8\textwidth]{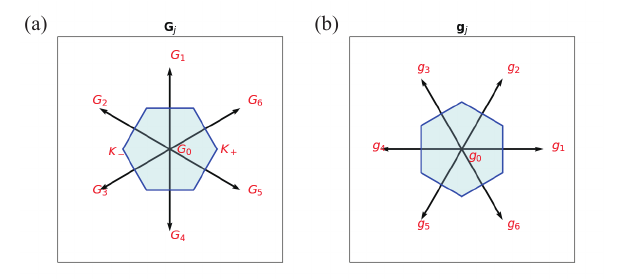}
    \caption{\label{fig:star} (a) The reciprocal lattice vectors of untwisted system $\pmb{G}_{0} = \mathbf{0}$, $\pmb{G}_{1} = \frac{4\pi}{\sqrt{3}a_{0}}\hat{\pmb{e}}_{y}$, and other $\pmb{G}_{j}$ are obtained by rotating $\pmb{G}_{1}$ by $\frac{j-1}{3}\pi$. (b) The reciprocal lattice vectors of the twisted system, the subscript number corresponds to $\pmb{G}_{j}$.}
\end{figure}

Next, consider the interlayer binding energy $U_B$. For bilayers like graphene or transition metal dichalcogenides (TMDs), the binding energy density can be modeled as \cite{lebedeva_interlayer_2011,verhoeven_model_2004}
\begin{equation}
    L_{B}(d_{z}, \pmb{\delta}) = V_{1}(d_{z}) + \sum_{j \in \{1,3,5\}} 2V_{B}V_{0}(d_{z}) \cos(\pmb{G}_{j} \cdot \pmb{\delta}).
\end{equation}
Here, $\pmb{\delta}$ is the in-plane interlayer displacement relative to AA stacking, determining the stacking configuration. The vectors $\pmb{G}_{j}$ are reciprocal lattice vectors: $\pmb{G}_{0} = \mathbf{0}$, $\pmb{G}_{1} = \frac{4\pi}{\sqrt{3}a_{0}}\hat{\pmb{e}}_{y}$, and other $\pmb{G}_{j}$ are obtained by rotating $\pmb{G}_{1}$ by $\frac{j-1}{3}\pi$, as shown in Fig.~\ref{fig:star}(a). Here, $a_{0}$ is the lattice constant of the untwisted bilayer, $V_{B}$ is a material-dependent binding energy coefficient, and $d_{z}$ is the interlayer distance.

While the exact forms of $V_{1}(d_{z})$ and $V_{0}(d_{z})$ are material-specific, their basic physical behavior is known. $V_{1}(d_{z})$ depends solely on integrals spacing $d_z$ and incorporates both van der Waals attraction and repulsion, possessing a minimum at an equilibrium spacing. $V_{1}(d_{z})$ is modeled as
\begin{equation}
    V_{1}(d_{z}) = 4\xi \left[ \exp\left(-\frac{d_{z} - c_{1}}{\beta}\right) - \exp\left(-\frac{d_{z} - c_{2}}{\beta}\right) \right],
\end{equation}
where $c_{2} > c_{1}$ are constants close to the equilibrium spacing, and $\beta$ is a decay constant. $V_{0}(d_{z})$ decays with increasing $d_z$ because the influence of stacking configuration on binding energy diminishes rapidly. Thus,
\begin{equation}
    V_{0}(d_{z}) = \epsilon \exp\left(-\frac{d_{z} - c_{0}}{\alpha}\right),
\end{equation}
here, $c_{0}$ is the minimum interlayer distance achievable across all stacking configurations (e.g., $c_{0} = d_{z}^{AB}$ for AB stacking). While $d_z$ is arbitrary, the equilibrium interlayer distance $d_{z}^{-}(\pmb{\delta}) = d_{z}^{(t)} - d_{z}^{(b)}$ at position $\pmb{r}$ is found by minimizing the total energy. The binding energy density then becomes
\begin{equation}
    L_{B}(\pmb{\delta}) = V_{1}(d_{z}^{-}) + \sum_{j \in \{1,3,5\}} 2V_{B}V_{0}(d_{z}^{-}) \cos(\pmb{G}_{j} \cdot \pmb{\delta}).
\end{equation}

Adapting the binding energy to the twisted case, and using the definition of $\pmb{u}^{-}$ (where $d_{z}^{-}$ is the equilibrium spacing), we get
\begin{equation}\label{eq:binding_energy}
    L_{B}(\pmb{r}) = 4\xi \left[ \exp\left(-\frac{d_{z}^{-} - c_{1}}{\beta}\right) - \exp\left(-\frac{d_{z}^{-} - c_{2}}{\beta}\right) \right] +
    \sum_{j \in \{1,3,5 \}} 2V_{B}\epsilon \exp\left(-\frac{d_{z}^{-} - c_{0}}{\alpha}\right)
    \cos\left[\pmb{g}_{j} \cdot \pmb{r} + \pmb{G}_{j} \cdot \pmb{u}^{-}\right].
\end{equation}
Here, we applied the local stacking approximation with the key relations
\begin{equation}
    \pmb{\delta} = \pmb{d}(\theta) + \pmb{u}^{-}_{\parallel},
\end{equation}
\begin{equation}
    \pmb{G}_{j} \cdot \pmb{d}(\theta) = \pmb{g}_{j} \cdot \pmb{r},
\end{equation}
where $\pmb{d}(\theta) = [R(\theta/2) - R(-\theta/2)] \cdot \pmb{r}$ is the local interlayer displacement due solely to the twist angle $\theta$, $R(\theta)$ is the rotation matrix, and $\pmb{g}_{j} = [R(-\theta/2) - R(\theta/2)] \cdot \pmb{G}_{j}$ are the moiré reciprocal lattice vectors at twist angle $\theta$. Note that since the z-component of $\pmb{G}_{j}$ is zero, there is $\pmb{G}_{j} \cdot \pmb{u}^{-}=\pmb{G}_{j} \cdot \pmb{u}^{-}_{\parallel}$. Consequently, $d_{z}^{-}(\pmb{\delta}) \to d_{z}^{-}(\pmb{r})$ and $\pmb{G}_{j} \cdot \pmb{\delta} \to \pmb{g}_{j} \cdot \pmb{r} + \pmb{G}_{j} \cdot \pmb{u}^{-}$. Here,
\begin{equation}
    d_{z}^{-}(\pmb{r}) = d_{0} + u_{z}^{-}(\pmb{r}),
\end{equation}
represents the local equilibrium interlayer distance at position $\pmb{r} = (x, y)$ in the twisted bilayer. $d_{0}$ is a constant representing the interlayer distance in the rigid (unrelaxed) case, and $u_{z}^{-}(\pmb{r})$ is the magnitude of the out-of-plane component defined in Eq.~(\ref{eq:upm}), signifying the deviation of the equilibrium spacing from the rigid case, with $u_{z}^{-}(\pmb{r}) \ll d_{0}$.

The total interlayer binding energy is thus
\begin{equation}
    U_{B} = \int \left\{
    4\xi \left[ \exp\left(-\frac{d_{z}^{-} - c_{1}}{\beta}\right) - \exp\left(-\frac{d_{z}^{-} - c_{2}}{\beta}\right) \right] +
    \sum_{j \in \{1,3,5 \}} 2V_{B}\epsilon \exp\left(-\frac{d_{z}^{-} - c_{0}}{\alpha}\right)
    \cos\left[\pmb{g}_{j} \cdot \pmb{r} + \pmb{G}_{j} \cdot \pmb{u}^{-}\right] \right\} d \pmb{r}.
\end{equation}
We reiterate that the parameters $c_{0}, c_{1}, c_{2}, \alpha, \beta$ are all on the order of the equilibrium interlayer distance $d_{z}^{-}$.

From the Euler-Lagrange equations Eq.~(\ref{eq:Lup}), we find for $\pmb{u}^{+}$
\begin{equation}
    \begin{aligned}
        &\frac{\partial^{2}u^{+}_{x}}{\partial x^{2}} + \frac{\partial^{2}u^{+}_{x}}{\partial y^{2}} = 0, \\
        &\frac{\partial^{2}u^{+}_{y}}{\partial x^{2}} + \frac{\partial^{2}u^{+}_{y}}{\partial y^{2}} = 0, \\
        &\frac{\partial^{2}u^{+}_{z}}{\partial x^{2}} + \frac{\partial^{2}u^{+}_{z}}{\partial y^{2}} = 0. \\
    \end{aligned}
\end{equation}
Since $\pmb{u}^{+}$ must be bounded over the entire 2D plane and vanishes at the origin, by Liouville's Theorem in complex analysis, we can directly obtain $\pmb{u}^{+} \equiv \mathbf{0}$. This implies $\pmb{u}^{(t)}(\pmb{r}) = -\pmb{u}^{(b)}(\pmb{r})$ in the 2D whole space.

The equations for $\pmb{u}^{-}$ Eq.~(\ref{eq:Lun}) become
\begin{equation}\label{eq:u_n}
    \begin{aligned}
        &\frac{\partial^{2}u^{-}_{x}}{\partial x^{2}} + \frac{\partial^{2}u^{-}_{x}}{\partial y^{2}} =
        \frac{2}{\mu}\frac{\partial L}{\partial u_{x}^{-}}, \\
        &\frac{\partial^{2}u^{-}_{y}}{\partial x^{2}} + \frac{\partial^{2}u^{-}_{y}}{\partial y^{2}} =
        \frac{2}{\mu}\frac{\partial L}{\partial u_{y}^{-}}, \\
        &\frac{\partial^{2}u^{-}_{z}}{\partial x^{2}} + \frac{\partial^{2}u^{-}_{z}}{\partial y^{2}} =
        \frac{1}{\zeta}\frac{\partial L}{\partial u_{z}^{-}}. \\
    \end{aligned}
\end{equation}
The first two equations govern in-plane relaxation, while the third governs out-of-plane relaxation.

\section{ Analytic Solutions for the Relaxation Equations}
Eq.~(\ref{eq:u_n}) lacks exact analytic solutions due to its complexity and coupling between components. To facilitate further analysis, we derive quantitatively accurate analytic solutions. We treat in-plane and out-of-plane relaxation separately.

\subsection{In-plane Relaxation}
For in-plane relaxation (first two equations of Eq.~(\ref{eq:u_n}))
\begin{equation}
    \begin{aligned}
        &\frac{\partial^{2}u^{-}_{x}}{\partial x^{2}} + \frac{\partial^{2}u^{-}_{x}}{\partial y^{2}} =
        -\sum_{j \in \{1,3,5\}} \kappa_{\parallel} \exp\left(-\frac{d_{z}^{-} - c_{0}}{\alpha}\right)
        \sin\left[\pmb{g}_{j} \cdot \pmb{r} + \pmb{G}_{j} \cdot \pmb{u}^{-}\right] G_{jx}, \\
        &\frac{\partial^{2}u^{-}_{y}}{\partial x^{2}} + \frac{\partial^{2}u^{-}_{y}}{\partial y^{2}} =
        -\sum_{j \in \{1,3,5\}} \kappa_{\parallel} \exp\left(-\frac{d_{z}^{-} - c_{0}}{\alpha}\right)
        \sin\left[\pmb{g}_{j} \cdot \pmb{r} + \pmb{G}_{j} \cdot \pmb{u}^{-}\right] G_{jy}. \\
    \end{aligned}
\end{equation}
Here we define the dimensionless coefficient $\kappa_{\parallel}=\frac{4V_B\epsilon}{\mu}$. We make two approximations, first, since $d_{z}^{-} - c_{0}$ is small compared to $\alpha$, we neglect the coupling between in-plane and out-of-plane relaxation, setting $\exp\left(-\frac{d_{z}^{-} - c_{0}}{\alpha}\right) \approx 1$. Second, for not-too-small twist angles, in-plane relaxation is small, so we neglect the self-coupling term $\pmb{G}_{j} \cdot \pmb{u}^{-}$ in the sine arguments. The validity of these approximations is verified numerically in the next section.

The simplified equations are
\begin{equation}
    \begin{aligned}
        &\frac{\partial^{2}u^{-}_{x}}{\partial x^{2}} + \frac{\partial^{2}u^{-}_{x}}{\partial y^{2}} =
        -\sum_{j \in \{1,3,5\}} \kappa_{\parallel} \sin(\pmb{g}_{j} \cdot \pmb{r}) G_{jx}, \\
        &\frac{\partial^{2}u^{-}_{y}}{\partial x^{2}} + \frac{\partial^{2}u^{-}_{y}}{\partial y^{2}} =
        -\sum_{j \in \{1,3,5\}} \kappa_{\parallel} \sin(\pmb{g}_{j} \cdot \pmb{r}) G_{jy}. \\
    \end{aligned}
\end{equation}

Exploiting the moiré periodicity of $\pmb{u}(\pmb{r})$, we perform a Fourier transform
\begin{equation}
    \pmb{u}^{-}_{\parallel}(\pmb{r}) = \sum_{i} \tilde{\pmb{u}}^{-}_{i} e^{i \pmb{g}_{i} \cdot \pmb{r}}.
\end{equation}
Here, tildes denote Fourier coefficients. The left-hand sides become
\begin{equation}
    \begin{aligned}
        &-\sum_{i} (g_{ix}^{2} \tilde{u}_{ix}^{-} + g_{iy}^{2} \tilde{u}_{ix}^{-}) e^{i \pmb{g}_{i} \cdot \pmb{r}} =
        -\sum_{i} |\pmb{g}_{i}|^{2} \tilde{u}_{ix}^{-} e^{i \pmb{g}_{i} \cdot \pmb{r}}, \\
        &-\sum_{i} (g_{ix}^{2} \tilde{u}_{iy}^{-} + g_{iy}^{2} \tilde{u}_{iy}^{-}) e^{i \pmb{g}_{i} \cdot \pmb{r}} =
        -\sum_{i} |\pmb{g}_{i}|^{2} \tilde{u}_{iy}^{-} e^{i \pmb{g}_{i} \cdot \pmb{r}}. \\
    \end{aligned}
\end{equation}
The right-hand sides become
\begin{equation}
    \begin{aligned}
        &-\sum_{j \in \{1,3,5\}} \kappa_{\parallel} \sin(\pmb{g}_{j} \cdot \pmb{r}) G_{jx} =
        -\sum_{j \in \{1,3,5\}} \kappa_{\parallel} \frac{e^{i \pmb{g}_{j} \cdot \pmb{r}} - e^{-i \pmb{g}_{j} \cdot \pmb{r}}}{2i} G_{jx} = -\sum_{j=1}^{6} \frac{\kappa_{\parallel}}{2i} G_{jx} e^{i \pmb{g}_{j} \cdot \pmb{r}}, \\
        &-\sum_{j \in \{1,3,5\}} \kappa_{\parallel} \sin(\pmb{g}_{j} \cdot \pmb{r}) G_{jy} =
        -\sum_{j \in \{1,3,5\}} \kappa_{\parallel} \frac{e^{i \pmb{g}_{j} \cdot \pmb{r}} - e^{-i \pmb{g}_{j} \cdot \pmb{r}}}{2i} G_{jy} = -\sum_{j=1}^{6} \frac{\kappa_{\parallel}}{2i} G_{jy} e^{i \pmb{g}_{j} \cdot \pmb{r}}. \\
    \end{aligned}
\end{equation}
As shown in Fig.~(\ref{fig:star}),$\{\pmb{G}_{1},\pmb{G}_{3},\pmb{G}_{5}\}$ and $\{\pmb{G}_{4},\pmb{G}_{6},\pmb{G}_{2}\}$ are equal in magnitude but opposite in direction (the same applies to $\pmb{g}_{j}$). Therefore, in the last equality of the expressions above, the negative sign for factor $e^{-i \pmb{g}_{j} \cdot \pmb{r}}$ is incorporated into $\pmb{G}_{j}$. Equating coefficients for each $e^{i \pmb{g}_{i} \cdot \pmb{r}}$
\begin{equation}
    \begin{aligned}
        &\sum_{i} |\pmb{g}_{i}|^{2} \tilde{u}_{ix}^{-} e^{i \pmb{g}_{i} \cdot \pmb{r}} = \sum_{i=1}^{6} \frac{\kappa_{\parallel}}{2i} G_{ix} e^{i \pmb{g}_{i} \cdot \pmb{r}}, \\
        &\sum_{i} |\pmb{g}_{i}|^{2} \tilde{u}_{iy}^{-} e^{i \pmb{g}_{i} \cdot \pmb{r}} = \sum_{i=1}^{6} \frac{\kappa_{\parallel}}{2i} G_{iy} e^{i \pmb{g}_{i} \cdot \pmb{r}}. \\
    \end{aligned}
\end{equation}
Due to the arbitrariness of $\pmb{r}$, for each $i \in \{1\cdots6\}$
\begin{equation}
    \begin{aligned}
        &|\pmb{g}_{i}|^{2} \tilde{u}_{ix}^{-} = \frac{\kappa_{\parallel}}{2i} G_{ix},\\
        &|\pmb{g}_{i}|^{2} \tilde{u}_{iy}^{-} = \frac{\kappa_{\parallel}}{2i} G_{iy}.
    \end{aligned}
\end{equation}
Combining components,
\begin{equation}
    \tilde{\pmb{u}}_{i}^{-} = \frac{\kappa_{\parallel}}{2i} \frac{1}{|\pmb{g}_{i}|^{2}} \pmb{G}_{i} \approx \frac{\kappa_{\parallel}}{2i}\frac{1}{|\pmb{G}_{i}|} \frac{1}{\theta^{2}} \hat{\pmb{G}}_{i} = \frac{\kappa_{\parallel}}{2i}\frac{1}{|\pmb{G}_{i}|} \frac{1}{\theta^{2}} \hat{z} \times \hat{\pmb{g}}_{i}.
\end{equation}
Note that, since $\pmb{g}_{i}=[R(-\frac{\theta}{2})-R(\frac{\theta}{2})]\pmb{G}_{i}=2\sin(\frac{\theta}{2})\pmb{G}_{i}\times \pmb{e}_{z}$, when $\theta$ is not too large, $2\sin(\frac{\theta}{2})\approx \theta$ holds, therefore, we use $|\pmb{g}_{i}|\approx \theta|\pmb{G}_{i}|$ above. The final analytical form for the in-plane relaxation displacement is
\begin{equation}\label{eq:in_palne_analysis}
    \pmb{u}^{-}_{\parallel}(\pmb{r}) = \sum_{i \in \{1,3,5\}} \frac{\kappa_{\parallel}}{|\pmb{G}_{i}|} \frac{1}{\theta^{2}} \hat{\pmb{G}}_{i} \sin(\pmb{g}_{i} \cdot \pmb{r}).
\end{equation}
This shows that the amplitude of the in-plane relaxation field scales inversely with the square of the twist angle ($\theta^{-2}$). Smaller angles lead to more significant in-plane relaxation. The amplitude is determined by the ratio of the material-dependent binding strength $V_B$ to the $\mu$. This successfully separates the geometric effect ($\theta^{-2}$) from the intrinsic material property ($V_B/\mu$), providing a concise analytical framework for understanding lattice reconstruction in twisted systems. This expression agrees with DFT, molecular dynamics,  and experimental results\cite{ezzi_analytical_2024,uchida_atomic_2014,shabani_deep_2021,liuMoirePhononsMagicAngle2022,liuDPmoireToolConstructing2024}.

\subsection{ Out-of-plane Relaxation}
Starting from the binding energy density Eq.~(\ref{eq:binding_energy})
\begin{equation}\label{eq:binding_energy_out}
    L_{B}(\pmb{r}) = 4\xi \left[ \exp\left(-\frac{d_{z}^{-} - c_{1}}{\beta}\right) - \exp\left(-\frac{d_{z}^{-} - c_{2}}{\beta}\right) \right] +
    \sum_{j \in \{1,3,5 \}} 2V_{B}\epsilon \exp\left(-\frac{d_{z}^{-} - c_{0}}{\alpha}\right)
    \cos\left[\pmb{g}_{j} \cdot \pmb{r} + \pmb{G}_{j} \cdot \pmb{u}^{-}\right].
\end{equation}
For not-too-small twist angles, relaxation displacements are small fluctuations. As before, we neglect the coupling between in-plane and out-of-plane relaxation. Expanding the exponentials for small arguments ($e^{-x} \approx 1 - x$)
\begin{equation}
    \begin{aligned}
        \exp\left(-\frac{d_{z}^{-} - c_{1}}{\beta}\right) &\approx 1 - \frac{d_{z}^{-} - c_{1}}{\beta}, \\
        \exp\left(-\frac{d_{z}^{-} - c_{2}}{\beta}\right) &\approx 1 - \frac{d_{z}^{-} - c_{2}}{\beta}, \\
        \exp\left(-\frac{d_{z}^{-} - c_{0}}{\alpha}\right) &\approx 1 - \frac{d_{z}^{-} - c_{0}}{\alpha} = 1 - \frac{u_{z}^{-} + d_{0} - c_{0}}{\alpha}.
    \end{aligned}
\end{equation}
yielding
\begin{equation}
    L_{B} \approx \frac{4\xi(c_1 - c_2)}{\beta} + \sum_{j \in \{1,3,5 \}} 2V_{B}\epsilon \left(1 - \frac{u_{z}^{-} + d_{0} - c_{0}}{\alpha}\right) \cos(\pmb{g}_{j} \cdot \pmb{r}).
\end{equation}
Inserting into the Euler-Lagrange equation for $u_z^{-}$
\begin{equation}
    \frac{\partial^{2}u^{-}_{z}}{\partial x^{2}} + \frac{\partial^{2}u^{-}_{z}}{\partial y^{2}} = -\sum_{j \in \{1,3,5 \}} \kappa_{\perp} \cos(\pmb{g}_{j} \cdot \pmb{r}).
\end{equation}
Here $\kappa_{\perp}=\frac{2V_{B}\epsilon}{\zeta\alpha}$. Exploiting moiré periodicity, the Fourier transform is performed as
\begin{equation}
    u_{z}^{-} = \sum_{i} \tilde{u}_{zi}^{-} e^{i \pmb{g}_{i} \cdot \pmb{r}},
\end{equation}
yielding
\begin{equation}
    \sum_{i} |\pmb{g}_{i}|^{2} \tilde{u}_{zi}^{-} e^{i \pmb{g}_{i} \cdot \pmb{r}} = \sum_{j \in \{1,3,5 \}} \kappa_{\perp} \cos(\pmb{g}_{j} \cdot \pmb{r}) = \frac{\kappa_{\perp}}{2} \sum_{i=1}^{6} e^{i \pmb{g}_{i} \cdot \pmb{r}}.
\end{equation}
Equating coefficients
\begin{equation}
    \tilde{u}_{zi}^{-} = \frac{1}{|\pmb{g}_{i}|^{2}} \frac{\kappa_{\perp}}{2}.
\end{equation}
Thus, the out-of-plane relaxation field is
\begin{equation}
    \begin{aligned}
        u_{z}^{-}(\pmb{r}) &= \sum_{i \in \{1,3,5\}} \frac{\kappa_{\perp}}{|\pmb{g}_{i}|^{2}} \cos(\pmb{g}_{i} \cdot \pmb{r}) \\
        &\approx \frac{1}{\theta^{2}} \kappa_{\perp} \sum_{i \in \{1,3,5\}} \frac{1}{|\pmb{G}_{i}|^{2}} \cos(\pmb{g}_{i} \cdot \pmb{r}).
    \end{aligned}
\end{equation}
The local interlayer distance is
\begin{equation}\label{eq:out_plane_analysis}
    \begin{aligned}
        d_{z}^{-}(\pmb{r}) &= d_{0} + u_{z}^{-}(\pmb{r}) \\
        &= d_{0} + u_{\perp}^{-}(\pmb{r}) \\
        &= d_{0} + \frac{1}{\theta^{2}} \kappa_{\perp} \sum_{i \in \{1,3,5\}} \frac{1}{|\pmb{G}_{i}|^{2}} \cos(\pmb{g}_{i} \cdot \pmb{r}).
    \end{aligned}
\end{equation}
The cosine terms, constrained by $C_3$ symmetry and moiré periodicity, ensure maximum interlayer distance at AA regions and minimum at AB regions, consistent with the prior results. The $\theta^{-2}$ factor causes pronounced layer distance modulation at small angles. This expression quantitatively describes the spatial modulation of interlayer spacing in twisted bilayers, revealing the physical origin of the moiré-induced "corrugation" effect.

\subsection{ Range of Validity for Analytic Solutions}
Our analytic solutions for in-plane (Eq.~\ref{eq:in_palne_analysis}) and out-of-plane (Eq.~\ref{eq:out_plane_analysis}) relaxation are
\begin{equation}
    \begin{aligned}
        \pmb{u}^{-}_{\parallel}(\pmb{r}) &= \frac{1}{\theta^{2}} \kappa_{\parallel} \sum_{i \in \{1,3,5\}} \frac{1}{|\pmb{G}_{i}|} \hat{z} \times \hat{\pmb{g}}_{i} \sin(\pmb{g}_{i} \cdot \pmb{r}), \\
        d_{z}^{-}(\pmb{r}) &= d_{0} + \frac{1}{\theta^{2}} \kappa_{\perp} \sum_{i \in \{1,3,5\}} \frac{1}{|\pmb{G}_{i}|^{2}} \cos(\pmb{g}_{i} \cdot \pmb{r}).
    \end{aligned}
\end{equation}
Both amplitudes diverge as $\theta \to 0$, which is unphysical. This stems from neglecting the $\pmb{G}_{j} \cdot \pmb{u}^{-}$ term in the derivations, which becomes non-ignorable at very small angles. We now establish the validity range.

For out-of-plane relaxation, the maximum ($d_{AA}$) and minimum ($d_{AB}$) interlayer spacing in the twisted bilayer approach those of the untwisted bilayer ($d_{AA}^{\text{untwisted}}$, $d_{AB}^{\text{untwisted}}$). Thus, the modulation amplitude should be bounded
\begin{equation}
    \frac{1}{\theta^{2}} \kappa_{\perp} \left[ \max_{\pmb{r}} \left( \sum_{i \in \{1,3,5\}} \frac{1}{|\pmb{G}_{i}|^{2}} \cos(\pmb{g}_{i} \cdot \pmb{r}) \right) - \min_{\pmb{r}} \left( \sum_{i \in \{1,3,5\}} \frac{1}{|\pmb{G}_{i}|^{2}} \cos(\pmb{g}_{i} \cdot \pmb{r}) \right) \right] \leq d_{AA} - d_{AB}.
\end{equation}
All $|\pmb{G}_{i}|$ in the first shell, as shown in Fig.~\ref{fig:star} are equal ($|\pmb{G}_{1}|$), The maximum and minimum values here can be directly calculated, thus
\begin{equation}
    \frac{1}{\theta^{2}} \kappa_{\perp} \frac{1}{|\pmb{G}_{1}|^{2}} \times 4.5 \leq d_{AA} - d_{AB}.
\end{equation}
Solving for $\theta$:
\begin{equation} 
    \theta \geq \frac{3}{|\pmb{G}_{1}|} \sqrt{\frac{V_B \epsilon}{\zeta \alpha} \frac{1}{d_{AA} - d_{AB}}} \equiv \theta^{*},
\end{equation}
This condition $\theta \geq \theta^{*}$ defines the validity range for both analytic solutions, as the same neglect of coupling was used in their derivation.

\section{ Numerical Solution of the Relaxation Equations}

Before proceeding to the numerical solution, we discuss the values of the relaxation parameters $\kappa_{\parallel}$ and $\kappa_{\perp}$. It should be noted that multiple factors influence these parameters, and there exist discrepancies in experimental measurements and DFT calculations. Therefore, in this work, we treat them as adjustable parameters, and for actual materials, we select reasonable values within the range of generally accepted results. In fact, their values are all very small, and within a reasonable order of magnitude, our calculation results will not differ much.

Regarding $\kappa_{\parallel} = \frac{4V_{B}\epsilon}{\mu}$, this dimensionless parameter characterizes the ratio between interlayer interactions and intralayer in-plane elastic deformation. For graphene, $\kappa_{\parallel} \approx 10^{-4}$\cite{san-jose_spontaneous_2014,popov_commensurateincommensurate_2011,lebedeva_interlayer_2011,jung_origin_2015,zakharchenko_finite_2009,nam_lattice_2017}. For TMDs, $\kappa_{\parallel} \approx 5 \times 10^{-4}$\cite{enaldievStackingDomainsDislocation2020,iguiniz_revisiting_2019,androulidakis_tailoring_2018}. For $\kappa_{\perp} = \frac{2V_{B}\epsilon}{\zeta \alpha}$. It describes the ratio between interlayer interaction and intralayer out-of-plane elastic deformation,which has units of $\si{\angstrom}^{-1}$, Both graphene and TMDs exhibit $\kappa_{\perp} \approx 10^{-4}$ $\si{\angstrom}^{-1}$\cite{iguiniz_revisiting_2019,androulidakis_tailoring_2018,shenoy_edgestressinduced_2008,wei_bending_2013,hajgato_outofplane_2013,calis_blister_2025,teklu_experimental_2024,fasolino_intrinsic_2007}.

Back to the relaxation equations, the full equations are
\begin{equation}
    \begin{aligned}
    \frac{\partial^{2}u^{-}_{x}}{\partial x^{2}} + \frac{\partial^{2}u^{-}_{x}}{\partial y^{2}} &=
    -\sum_{j \in \{1,3,5\}} \kappa_{\parallel} \sin\left[\pmb{g}_{j} \cdot \pmb{r} + \pmb{G}_{j} \cdot \pmb{u}^{-}\right] G_{jx}, \\
    \frac{\partial^{2}u^{-}_{y}}{\partial x^{2}} + \frac{\partial^{2}u^{-}_{y}}{\partial y^{2}} &=
    -\sum_{j \in \{1,3,5\}} \kappa_{\parallel} \sin\left[\pmb{g}_{j} \cdot \pmb{r} + \pmb{G}_{j} \cdot \pmb{u}^{-}\right] G_{jy}, \\
    \frac{\partial^{2}u^{-}_{z}}{\partial x^{2}} + \frac{\partial^{2}u^{-}_{z}}{\partial y^{2}} &= -\sum_{j \in \{1,3,5 \}} \kappa_{\perp} \cos\left[\pmb{g}_{j} \cdot \pmb{r} + \pmb{G}_{j} \cdot \pmb{u}^{-}\right].
    \end{aligned}
\end{equation}
We retain the approximation $\exp\left(-\frac{d_{z}^{-} - c_{0}}{\alpha}\right) \approx 1$ in the first two equations, as its impact is negligible based on our calculations.

\subsection{ In-plane Relaxation}
Utilizing moiré periodicity, we Fourier transform the in-plane field
\begin{equation}
    \pmb{u}^{-}_{\parallel}(\pmb{r}) = \sum_{i} \tilde{\pmb{u}}^{-}_{i} e^{i \pmb{g}_{i} \cdot \pmb{r}}.
\end{equation}
The sine term $\sin (\pmb{g}_{j} \cdot \pmb{r} + \pmb{G}_{j} \cdot \pmb{u}^{-})$ is also periodic and can be expanded as
\begin{equation}
    \sin (\pmb{g}_{j} \cdot \pmb{r} + \pmb{G}_{j} \cdot \pmb{u}^{-}) = \sum_{i} \tilde{f}_{i}^{j} e^{i \pmb{g}_{i} \cdot \pmb{r}}.
\end{equation}
The equations become
\begin{equation}
    \begin{aligned}
        \sum_{i} |\pmb{g}_{i}|^{2} \tilde{u}_{ix}^{-} e^{i \pmb{g}_{i} \cdot \pmb{r}} &= \kappa_{\parallel} \sum_{i} \left[ \sum_{j \in \{1,3,5\}} \tilde{f}_{i}^{j} G_{jx} \right] e^{i \pmb{g}_{i} \cdot \pmb{r}}, \\
        \sum_{i} |\pmb{g}_{i}|^{2} \tilde{u}_{iy}^{-} e^{i \pmb{g}_{i} \cdot \pmb{r}} &= \kappa_{\parallel} \sum_{i} \left[ \sum_{j \in \{1,3,5\}} \tilde{f}_{i}^{j} G_{jy} \right] e^{i \pmb{g}_{i} \cdot \pmb{r}}. \\
    \end{aligned}
\end{equation}
Equating coefficients
\begin{equation}
    \tilde{\pmb{u}}_{i}^{-} = \kappa_{\parallel} \sum_{j \in \{1,3,5\}} \frac{\tilde{f}_{i}^{j}}{|\pmb{g}_{i}|^{2}} \pmb{G}_{j}.
\end{equation}
The self-coupling $\pmb{G}_{j} \cdot \pmb{u}^{-}$ inside $\tilde{f}_{i}^{j}$ makes this equation implicit. To solve it, we use an iterative approach
\begin{equation}
    \Delta \pmb{u}^{-,(n)}_{\parallel}(\pmb{r}) = -\sum_{j \in \{1,3,5\}} \kappa_{\parallel} \sin \left[\pmb{g}_{j} \cdot \pmb{r} + \frac{n}{N} \pmb{G}_{j} \cdot \pmb{u}^{-,(n-1)}(\pmb{r})\right] \pmb{G}_{j},
\end{equation}
for $n = 1, 2, \dots, N$, where $\Delta = \partial_x^2 + \partial_y^2$. The initial guess $\pmb{u}^{-,(0)}_{\parallel}(\pmb{r})$ is the analytic solution Eq.~(\ref{eq:in_palne_analysis}). The Fourier coefficients at step $n$ are
\begin{equation}
    \tilde{\pmb{u}}_{i}^{-,(n)} = \kappa_{\parallel} \sum_{j \in \{1,3,5\}} \frac{\tilde{f}_{i}^{j,(n)}}{|\pmb{g}_{i}|^{2}} \pmb{G}_{j},
\end{equation}
with
\begin{equation}
    \tilde{f}_{i}^{j,(n)} = \frac{1}{S_{0}} \int \sin \left[\pmb{g}_{j} \cdot \pmb{r} + \frac{n}{N} \pmb{G}_{j} \cdot \pmb{u}^{-,(n-1)}(\pmb{r})\right] e^{-i \pmb{g}_{i} \cdot \pmb{r}} \, d\pmb{r}.
\end{equation}
For sufficiently large $N$, this iteration converges to the numerical solution starting from the analytic solution.

\begin{figure}[t]
    \centering
    \includegraphics[width=1\textwidth]{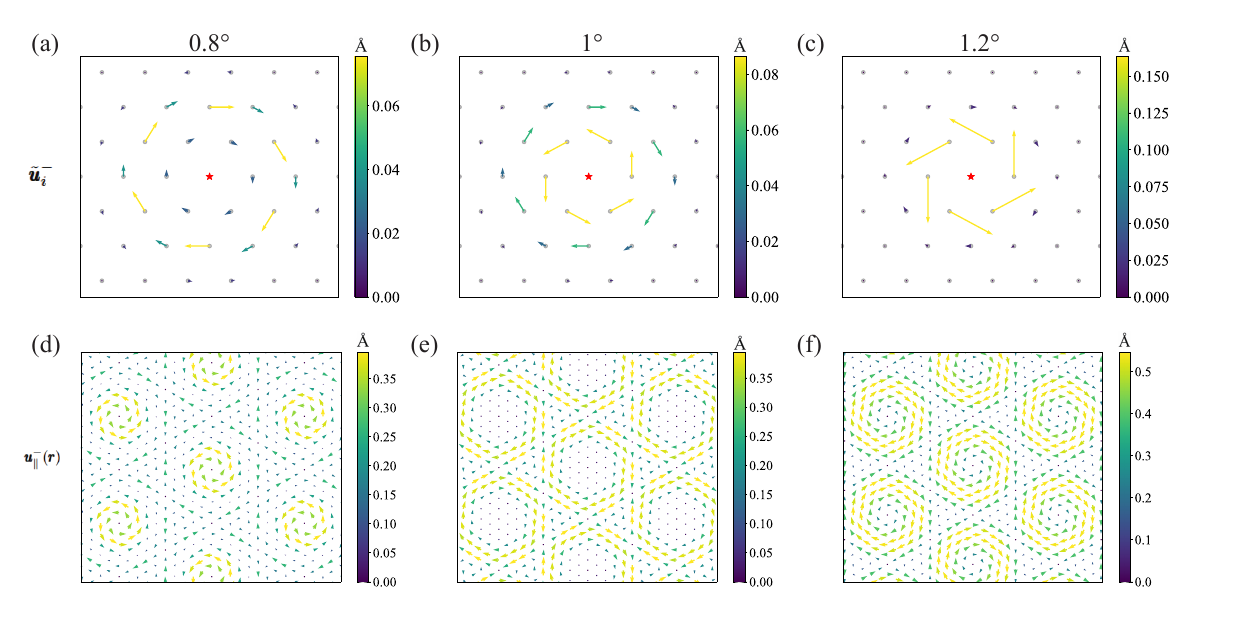}
    \caption{\label{fig:numerical} Numerical solutions of in-plane relaxation in tTMD at various twist angles. (a)-(c) Fourier component distributions of $\pmb{u}^{-}_{\parallel}(\pmb{r})$ in reciprocal space. (d)-(f) Real-space profiles of $\pmb{u}^{-}_{\parallel}(\pmb{r})$.}
\end{figure}

Below, taking tTMD as an example, we perform numerical solutions of the system, obtaining the results shown in Fig.~\ref{fig:numerical}. In-plane relaxation manifests as an additional rotation of atoms around the moiré period origin. Since only the Fourier components of the first star exist essentially in Fig. ~\ref{fig:numerical} (c), our analytic solution is already sufficiently accurate at $\theta=1.2^{\circ}$.

\subsection{ Out-of-plane Relaxation}
Once the in-plane field $\pmb{u}^{-}_{\parallel}(\pmb{r})$ is known numerically, the equation for out-of-plane relaxation is
\begin{equation}
    \frac{\partial^{2}u^{-}_{z}}{\partial x^{2}} + \frac{\partial^{2}u^{-}_{z}}{\partial y^{2}} = -\sum_{j \in \{1,3,5 \}} \kappa_{\perp} \cos\left[\pmb{g}_{j} \cdot \pmb{r} + \pmb{G}_{j} \cdot \pmb{u}^{-}\right].
\end{equation}
Fourier transforming both sides
\begin{equation}
    u_{z}^{-} = \sum_{i} \tilde{u}_{zi}^{-} e^{i \pmb{g}_{i} \cdot \pmb{r}},
\end{equation}
\begin{equation}
    \cos \left[\pmb{g}_{j} \cdot \pmb{r} + \pmb{G}_{j} \cdot \pmb{u}^{-}\right] = \sum_{i} \tilde{p}_{i}^{j} e^{i \pmb{g}_{i} \cdot \pmb{r}}.
\end{equation}
The coefficients $\tilde{p}_{i}^{j}$ are computed using the known $\pmb{u}^{-}_{\parallel}(\pmb{r})$. The solution is then
\begin{equation}
    \tilde{u}_{zi}^{-} = \kappa_{\perp} \sum_{j \in \{1,3,5\}} \frac{\tilde{p}_{i}^{j}}{|\pmb{g}_{i}|^{2}}.
\end{equation}

\section{Impact of Relaxation on Electronic Structure}
Having derived expressions for in-plane and out-of-plane relaxation, we now discuss their effects on the electronic structure.

\subsection{ Effect of Out-of-plane Relaxation}
\subsubsection{ Interlayer Coupling}
Taking a hexagonal system as an example, consider an untwisted bilayer structure. Assume the bottom layer is fixed while the top layer undergoes displacement. Starting from the Wannier functions $|\pmb{R}^{t}_{X,i}\rangle$, $|\pmb{R}^{b}_{X,i}\rangle$ for an orbital $X$ within the monolayer, the positions of their Wannier centers can be written as
\begin{equation}
    \begin{aligned}
        \pmb{R}^{t}_{X,i} &= \pmb{R}_{i} + \pmb{\tau}_{X} + \pmb{\delta} +
        d_{z}^{-}(\pmb{\delta})\hat{\pmb{e}}_{z},\\
        \pmb{R}^{b}_{X,i} &= \pmb{R}_{i} + \pmb{\tau}_{X},
    \end{aligned}
\end{equation}
here, $l=\{t,b\}$ denotes the top and bottom layers, $\pmb{\tau}_X$ represents the position of the $X$-th orbital within the unit cell, $\pmb{\delta}$ is the in-plane relative displacement between the layers, which determines the stacking configuration of the bilayer. $\pmb{R}_{i}=n_{1,i}\pmb{a}_1+n_{2,i}\pmb{a}_2$, where $n_{1,i}, n_{2,i}$ are arbitrary integers, $\pmb{a}_1, \pmb{a}_2$ are lattice vectors. The term $d_{z}^{-}(\pmb{\delta}) \hat{\pmb{e}}_{z}$ is the relaxed interlayer distance, determined by $\pmb{\delta}$. Since we ultimately aim to calculate out-of-plane relaxation in twisted systems (where the relaxation magnitude depends on the twist angle $\theta$), which differs from the untwisted bilayer case, we adopt the form from Eq. (\ref{eq:out_plane_analysis}) for not-too-small ($\theta > \theta^{*}$) angles and transform it via the local stacking approximation,
\begin{equation}
    d_{z}^{-}(\pmb{\delta})=d_{0}+\frac{1}{[\max(\theta,\theta^{*})]^{2}}\frac{2V_{B}\epsilon}{\zeta\alpha} \frac{1}{|\pmb{G}_{i}|^{2}}\sum_{i \in \{1,3,5\}}
    \cos(\pmb{G}_{i}\cdot \pmb{\delta}).
\end{equation}
It should be noted that above, we have used the relation $\pmb{G}_{i} \cdot \pmb{\delta}=\pmb{g}_{i} \cdot \pmb{r}$, which means the influence of in-plane relaxation on out-of-plane relaxation is neglected, as this influence is small. The factor $\frac{1}{[\max(\theta,\theta^{*})]^{2}}$ accounts for the effect of the twist angle on relaxation at general angles while avoiding divergence at very small angles.

On the other hand, we can construct the Bloch functions in k-space from the Wannier functions
\begin{equation}
|\pmb{k},l,X\rangle=\frac{1}{\sqrt{N}}\sum_{i}e^{i\pmb{k}\cdot(\pmb{R}_{i} + \pmb{\tau}_{X})} 
|\pmb{R}^{l}_{X,i}\rangle.
\end{equation}
For the weak, distance-dependent van der Waals interlayer interaction, we use the two-center approximation,
\begin{equation}
\begin{aligned}
\langle \pmb{R}^{t}_{X',i'}|H|\pmb{R}^{b}_{X,i}\rangle
&=-t(\pmb{R}^{t}_{X',i'}-\pmb{R}^{b}_{X,i})\\
&=-t(\pmb{R}_{i'i}+\pmb{\tau}_{X'X}+\pmb{\delta}+d_{z}^{-}(\pmb{\delta})\hat{\pmb{e}}_{z}).
\end{aligned}
\end{equation}
Here, we use the shorthand notations $\pmb{R}_{i'i}=\pmb{R}_{i'}-\pmb{R}_{i}$ and $\pmb{\tau}_{X'X}=\pmb{\tau}_{X'}-\pmb{\tau}_{X}$. The value of $t$ depends only on the distance between the two Wannier centers.

The interlayer coupling in momentum space is then
\begin{equation}
\begin{aligned}
T_{X'X}(\pmb{k},\pmb{\delta})
&=\langle \pmb{k},t,X'|H|\pmb{k}, b, X\rangle \\
&=-\frac{1}{N}\sum_{i'i}
t(\pmb{R}_{i'i}+\pmb{\tau}_{X'X}+\pmb{\delta}+d_{z}^{-}(\pmb{\delta})\hat{\pmb{e}}_{z})
\exp[-i\pmb{k}\cdot(\pmb{R}_{i'i}+\pmb{\tau}_{X'X})],
\end{aligned}
\end{equation}
Replacing the summation index $\pmb{R}_{i'i}=\pmb{R}_{i'}-\pmb{R}_{i}$ with $\pmb{R}_{i}$ (effectively summing over lattice vectors) eliminates the factor $\frac{1}{N}$, yielding
\begin{equation}\label{eq:real_inter_couple}
T_{X'X}(\pmb{k},\pmb{\delta})
=-\sum_{i}t(\pmb{R}_{i}+\pmb{\tau}_{X'X}+\pmb{\delta}+d_{z}^{-}(\pmb{\delta})\hat{\pmb{e}}_{z})
\exp[-i\pmb{k}\cdot(\pmb{R}_{i}+\pmb{\tau}_{X'X})].
\end{equation}
Performing a Fourier transform
\begin{equation}
    t(\pmb{R}_{i}+\pmb{\tau}_{X'X}+\pmb{\delta}+d_{z}^{-}(\pmb{\delta})\hat{\pmb{e}}_{z})=\int_{\mathbb{R}^{2}}\frac{d\pmb{p}}{(2\pi)^{2}}
    t_{X'X}(\pmb{p})
    e^{i\pmb{p}\cdot (\pmb{R}_{i}+\pmb{\tau}_{X'X}+\pmb{\delta})}.
\end{equation}
we obtain
\begin{equation}
    \begin{aligned}
        T_{X'X}(\pmb{k},\pmb{\delta})
        &=- \int_{\mathbb{R}^{2}}\frac{d\pmb{p}}{(2\pi)^{2}} t_{X'X} (\pmb{p}) e^{i\pmb{p}\cdot \pmb{\delta}}e^{i(\pmb{p}-\pmb{k})\cdot\pmb{\tau}_{X'X}}\sum_{i}e^{i(\pmb{p}-\pmb{k})\cdot\pmb{R}_{i}}\\
        &= -\int_{\mathbb{R}^{2}}d\pmb{p}\frac{1}{S_{0}}  e^{i\pmb{p}\cdot \pmb{\delta}}e^{i(\pmb{p}-\pmb{k})\cdot\pmb{\tau}_{X'X}} t_{X'X} (\pmb{p}) \sum_{j}\delta[\pmb{p}-(\pmb{k}+\pmb{G}_{j})]   \\
        &=-\sum_{j}\frac{1}{S_{0}}t_{X'X} (\pmb{k}+\pmb{G}_{j})e^{i(\pmb{k}+\pmb{G}_{j})\cdot\pmb{\delta}}e^{i\pmb{G}_{j}\cdot\pmb{\tau}_{X'X}}. 
    \end{aligned}
\end{equation}
In the second equality, we used the identity $\sum_{i}e^{i(\pmb{p}-\pmb{k})\cdot\pmb{R}_{i}} = N \sum_{j}\delta_{\pmb{p}-\pmb{k}, \pmb{G}_{j}}$ and the relation $\delta_{\pmb{k},\pmb{k}'}=\frac{(2\pi)^{2}}{NS_{0}}\delta(\pmb{k}-\pmb{k}')$ for the Kronecker delta expressed in terms of the Dirac delta function in the thermodynamic limit, where $S_{0}$ is the area of the monolayer's unit cell.

The Fourier component $t_{X'X}(\pmb{p})$ is defined by
\begin{equation}
    \begin{aligned}
        t_{X'X}(\pmb{p})&=\int_{\mathbb{R}^{2}}d\pmb{\delta} e^{-i\pmb{p}\cdot(\pmb{R}_{i}+\pmb{\tau}_{X'X}+\pmb{\delta})}
        t(\pmb{R}_{i}+\pmb{\tau}_{X'X}+\pmb{\delta}+d_{z}^{-}(\pmb{\delta})\pmb{e}_{z})\\
        &=\int_{\mathbb{R}^{2}}d\pmb{\delta} e^{-i\pmb{p}\cdot\pmb{\delta}}
        t(\pmb{\delta}+d_{z}^{-}(\pmb{\delta}-\pmb{\tau}_{X'X})\pmb{e}_{z}),\\
    \end{aligned}
\end{equation}
above, because the integration is over the entire 2D space and the function $t$ is expected to decay rapidly, the specific value of $\pmb{R}_{i} + \pmb{\tau}_{X'X}$ becomes irrelevant for the integral over $\pmb{\delta}$ (it's a shift), and we use the periodic condition $d_{z}^{-}(\pmb{\delta}+\pmb{R}_{i})=d_{z}^{-}(\pmb{\delta})$. Note, $t_{X'X}(\pmb{p})$ depends only on $|\pmb{p}|$.

In continuum models, we are often interested in the region near a specific valley point $\pmb{K}_{\eta}$ in the monolayer Brillouin zone, where $\eta$ is the valley index. Since $t_{X'X}(\pmb{K}_{\eta}+\pmb{G}_{j})$ decays rapidly with increasing $|\pmb{K}_{\eta}+\pmb{G}_{j}|$, we only need to consider the nearest few terms (e.g., for $\eta=+$, $j \in \{0,2,3\}$, for $\eta=-$, $j \in \{0,5,6\}$). We define
\begin{equation}
    \begin{aligned}
        \omega_{X'X}&=-\frac{1}{S_{0}}\int_{\mathbb{R}^{2}}d\pmb{\delta}  e^{-i\pmb{K}_{\eta}\cdot\pmb{\delta}}
        t(\pmb{\delta}+d_{z}^{-}(\pmb{\delta}-\pmb{\tau}_{X'X})\hat{\pmb{e}}_{z})\\
        &=-\frac{1}{S_{0}}t_{X'X}(\pmb{K}_{\eta}).
    \end{aligned}
\end{equation}
Thus, we obtain the interlayer coupling as a function of displacement
\begin{equation}\label{eq:inter_couple}
    T_{X'X}(\pmb{\delta})=\sum_{j \in \{0,2,3\}}\omega_{X'X}e^{i\pmb{G}_{j}\cdot\pmb{\tau}_{X'X}}e^{i(\pmb{K}_{\eta}+\pmb{G}_{j})\cdot\pmb{\delta}}.
\end{equation}
When in-plane relaxation is neglected ($\pmb{u}^{-} = 0$), the relative displacement is purely due to the twist: $\pmb{\delta} = \pmb{d}(\theta) = [R(\theta/2)-R(-\theta/2)]\pmb{r}$. In this case, $\pmb{G}_{j}\cdot\pmb{\delta}=\pmb{G}_{j}\cdot\pmb{d}(\theta)=\pmb{g}_{j}\cdot\pmb{r}$ and $\pmb{K}_{\eta}\cdot\pmb{\delta}=\pmb{K}_{\eta}\cdot\pmb{d}(\theta)=\pmb{q}_{\eta}\cdot\pmb{r}$ (where $\pmb{q}_{\eta}=[R(-\frac{\theta}{2})-R(\frac{\theta}{2})]\pmb{K}_{\eta}$), yielding the standard form used in continuum models (incorporated the effect of out-of-plane relaxation)
\begin{equation}
    T_{X'X}(\pmb{r})=\sum_{j \in \{0,2,3\}}\omega_{X'X}e^{i\pmb{G}_{j}\cdot\pmb{\tau}_{X'X}}e^{i(\pmb{q}_{\eta}+\pmb{g}_{j})\cdot\pmb{r}}.
\end{equation}
Up to this point, we have incorporated the effect of out-of-plane relaxation into the interlayer coupling of the continuum model, and its influence is mainly reflected in the coupling coefficients $\omega_{X'X}$.

\subsubsection{ Intralayer Potential}
Similarly, consider the intralayer part Hamiltonian
\begin{equation}\label{eq:rhr_ham}
    \langle \pmb{R}^{l}_{X',i'} | H | \pmb{R}^{l}_{X,i} \rangle = h_{X'X}^{l}(\pmb{R}_{i'i}; \pmb{\delta}, d_{z}^{-}(\pmb{\delta})).
\end{equation}
For $(\pmb{R}_{i'i}=0, X'=X)$, this is the on-site energy; otherwise, it represents hopping. Periodicity requires
\begin{equation}
    h_{X'X}^{l}(\pmb{R}_{i'i}; \pmb{\delta} + \pmb{R}_{j}) = h_{X'X}^{l}(\pmb{R}_{i'i}; \pmb{\delta}).
\end{equation}
The intralayer Hamiltonian in momentum space is
\begin{equation}
    \langle\pmb{k}, l, X'| H | \pmb{k}, l, X\rangle = \frac{1}{N} \sum_{i', i} h^{l}_{X'X}(\pmb{R}_{i'i}; \pmb{\delta}, d_{z}^{-}(\pmb{\delta})) \exp\left[-i \pmb{k} \cdot (\pmb{R}^{l}_{X',i'} - \pmb{R}^{l}_{X,i})\right].
\end{equation}
Shifting indices ($\pmb{R}_{i'i} \to \pmb{R}_{i}$)
\begin{equation}
    \langle\pmb{k}, l, X'| H | \pmb{k}, l, X\rangle = \sum_{i} h^{l}_{X'X}(\pmb{R}_{i}; \pmb{\delta}, d_{z}^{-}(\pmb{\delta})) \exp\left[-i \pmb{k} \cdot (\pmb{R}_{i} + \pmb{\tau}_{X'X})\right].
\end{equation}
This is periodic in $\pmb{\delta}$. Fourier transforming as
\begin{equation}
    \langle\pmb{k}, l, X'| H | \pmb{k}, l, X\rangle = \sum_{j} \nu^{l,j}_{X'X}(\pmb{k}, \pmb{G}_{j}) \exp\left[i \pmb{G}_{j} \cdot \pmb{\delta}\right],
\end{equation}
where
\begin{equation}\label{eq:V_XX_inver_trans}
    \nu^{l,j}_{X'X}(\pmb{k}, \pmb{G}_{j}) = \frac{1}{S_0} \sum_{i} \exp\left[-i \pmb{k} \cdot (\pmb{R}_{i} + \pmb{\tau}_{X'X})\right] \int_{S_{0}} h^{l}_{X'X}(\pmb{R}_{i}; \pmb{\delta}, d_{z}^{-}(\pmb{\delta})) \exp\left[-i \pmb{G}_{j} \cdot \pmb{\delta}\right] d\pmb{\delta}.
\end{equation}
The $j=0$ component is the average intralayer Hamiltonian
\begin{equation}
    H^{0}_{X'X}(\pmb{k}) = \nu^{l,0}_{X'X}(\pmb{k}, 0) = \frac{1}{S_0} \sum_{i} \exp\left[-i \pmb{k} \cdot (\pmb{R}_{i} + \pmb{\tau}_{X'X})\right] \int_{S_{0}} h^{l}_{X'X}(\pmb{R}_{i}; \pmb{\delta}, d_{z}^{-}(\pmb{\delta})) d\pmb{\delta},
\end{equation}
which is the stacking average part and also the monolayer part of the Hamiltonian in the continuum model. Focusing on a specific valley $\pmb{K}_{\eta}$ and transforming to the moiré lattice, the intralayer Hamiltonian becomes
\begin{equation}
    \langle\pmb{k}, l, X'| H | \pmb{k}, l, X\rangle = H^{0}_{X'X}(\pmb{K}_{\eta}) + \sum_{j \neq 0} \nu^{l,j}_{X'X}(\pmb{K}_{\eta}, \pmb{G}_{j}) \exp\left[i \pmb{G}_{j} \cdot \pmb{\delta}\right].
\end{equation}
Define the moiré potential $V^{l}_{X'X}(\pmb{\delta}) = \sum_{j \neq 0} \nu^{l}_{X'X}(\pmb{K}_{\eta}, \pmb{G}_{j}) \exp\left[i \pmb{G}_{j} \cdot \pmb{\delta}\right]$. Taking only the non-ignorable first shell of reciprocal vectors
\begin{equation}\label{eq:intra_potential}
    V^{l}_{X'X}(\pmb{\delta}) = \sum_{j=1}^{6} \nu_{X'X}^{l,j} \exp\left[i \pmb{G}_{j} \cdot \pmb{\delta}\right],
\end{equation}
Here, $d_z^{-}(\delta)$ modulates the local potential via charge redistribution. The form of $h^{l}_{X'X}(\pmb{R}_{i}; \pmb{\delta}, d_{z}^{-}(\pmb{\delta}))$ can be obtained from DFT calculations of untwisted bilayers at various displacements $\pmb{\delta}$, combined with Wannier function projection \cite{wu_topological_2019}. In this section, we incorporated the effect of out-of-plane relaxation on the intralayer Hamiltonian.

\subsection{ Effect of In-plane Relaxation}
In-plane relaxation modifies the interlayer displacement
\begin{equation}
    \pmb{\delta} = [R(\theta/2) - R(-\theta/2)] \pmb{r} + \pmb{u}_{\parallel}^{-}(\pmb{r}) = \pmb{d}(\pmb{r}) + \pmb{u}^{-}_{\parallel}(\pmb{r}).
\end{equation}
Consequently, $\pmb{G}_{j} \cdot \pmb{\delta} = \pmb{g}_{j} \cdot \pmb{r} + \pmb{G}_{j} \cdot \pmb{u}^{-}_{\parallel}(\pmb{r})$. Therefore, the interlayer coupling Eq.~(\ref{eq:inter_couple}) and intralayer potential Eq.~(\ref{eq:intra_potential}) become
\begin{equation}
    \begin{aligned}
        T_{X'X}(\pmb{r}) &= \sum_{i \in \{0,2,3\}} \omega_{X'X} \exp\left[i \pmb{G}_i \cdot \pmb{\tau}_{X'X}\right] \exp\left[i \pmb{G}_i \cdot \pmb{\delta}\right] \exp\left[i \pmb{K}_{\eta} \cdot \pmb{\delta}\right] \\
        &= \sum_{i \in \{0,2,3\}} \omega_{X'X} \exp\left[i \pmb{G}_i \cdot \pmb{\tau}_{X'X}\right] \exp\left[i (\pmb{g}_i + \pmb{q}_{\eta}) \cdot \pmb{r}\right] \exp\left[i (\pmb{G}_i + \pmb{K}_{\eta}) \cdot \pmb{u}^{-}_{\parallel}\right], \\
    \end{aligned}
\end{equation}
\begin{equation}
    \begin{aligned}
        V_{X'X}^{l}(\pmb{r}) &= \sum_{i=1}^{6} \nu_{X'X}^{l,i} \exp\left[i \pmb{G}_i \cdot \pmb{\delta}\right] \\
        &= \sum_{i=1}^{6} \nu_{X'X}^{l,i} \exp\left[i \pmb{g}_i \cdot \pmb{r}\right] \exp\left[i \pmb{G}_i \cdot \pmb{u}^{-}_{\parallel}\right]. \\
    \end{aligned}
\end{equation}
Using the analytic solution for in-plane relaxation Eq.~(\ref{eq:in_palne_analysis}), we have
\begin{equation}
    i (\pmb{G}_i + \pmb{K}_{\eta}) \cdot \pmb{u}^{-}_{\parallel} = i \sum_{j \in \{1,3,5\}} \frac{\kappa_{\parallel}}{|\pmb{G}_{j}|} \frac{1}{\theta^{2}} (\pmb{G}_i + \pmb{K}_{\eta}) \cdot \hat{\pmb{G}}_{j} \sin(\pmb{g}_{j} \cdot \pmb{r}) = \sum_{j=1}^{6} \frac{\kappa_{\parallel}}{2} \frac{1}{\theta^{2}} \frac{(\pmb{G}_i + \pmb{K}_{\eta})}{|\pmb{G}_{j}|} \cdot \hat{\pmb{G}}_{j} e^{i \pmb{g}_{j} \cdot \pmb{r}},
\end{equation}
\begin{equation}
    i \pmb{G}_i \cdot \pmb{u}^{-}_{\parallel} = \sum_{j=1}^{6} \frac{\kappa_{\parallel}}{2} \frac{1}{\theta^{2}} \hat{\pmb{G}}_{i} \cdot \hat{\pmb{G}}_{j} e^{i \pmb{g}_{j} \cdot \pmb{r}},
\end{equation}
Defining coefficients
\begin{equation}
    \begin{aligned}
        \gamma_{ij}' &:= \frac{\kappa_{\parallel}}{2} \frac{1}{\theta^{2}} \frac{(\pmb{G}_i + \pmb{K}_{\eta})}{|\pmb{G}_{j}|} \cdot \hat{\pmb{G}}_{j}, \\
        \gamma_{ij} &:= \frac{\kappa_{\parallel}}{2} \frac{1}{\theta^{2}} \hat{\pmb{G}}_{i} \cdot \hat{\pmb{G}}_{j}. \\
    \end{aligned}
\end{equation}
We expand the exponential factors
\begin{equation}\label{eq:exband_Q}
    \begin{aligned}
        \exp\left[i (\pmb{G}_i + \pmb{K}_{\eta}) \cdot \pmb{u}^{-}_{\parallel}(\pmb{r})\right] &= \sum_{n=0}^{\infty} \frac{[i (\pmb{G}_i + \pmb{K}_{\eta}) \cdot \pmb{u}^{-}_{\parallel}]^n}{n!} \\
        &= 1 + \sum_{j=1}^{6} \gamma_{ij}' e^{i \pmb{g}_{j} \cdot \pmb{r}} + \sum_{j_{1},j_{2}=1}^{6} \frac{\gamma_{ij_{1}}' \gamma_{ij_{2}}'}{2} e^{i (\pmb{g}_{j_{1}} + \pmb{g}_{j_{2}}) \cdot \pmb{r}} + \cdots ,
    \end{aligned}
\end{equation}
\begin{equation}\label{eq:exband_G}
    \begin{aligned}
        \exp\left[i \pmb{G}_i \cdot \pmb{u}^{-}_{\parallel}(\pmb{r})\right] &= \sum_{n=0}^{\infty} \frac{(i \pmb{G}_i \cdot \pmb{u}^{-}_{\parallel})^n}{n!} \\
        &=  1 + \sum_{j=1}^{6} \gamma_{ij} e^{i \pmb{g}_{j} \cdot \pmb{r}} + \sum_{j_{1},j_{2}=1}^{6} \frac{\gamma_{ij_{1}} \gamma_{ij_{2}}}{2} e^{i (\pmb{g}_{j_{1}} + \pmb{g}_{j_{2}}) \cdot \pmb{r}} + \cdots .
    \end{aligned}
\end{equation}
Substituting into $T_{X'X}(\pmb{r})$ and $V_{X'X}^{l}(\pmb{r})$ gives expansions
\begin{equation}
    \begin{aligned}
        T_{X'X}(\pmb{r}) &= \sum_{n=0}^{\infty} T_{X'X}^{(n)}(\pmb{r}), \\
        V_{X'X}^{l}(\pmb{r}) &= \sum_{n=0}^{\infty} V_{X'X}^{l,(n)}(\pmb{r}),
    \end{aligned}
\end{equation}
where
\begin{equation}\label{eq:exband_inter}
    T_{X'X}^{(n)}(\pmb{r}) = \sum_{i \in \{0,2,3\}} \sum_{j_1, \dots, j_n=1}^{6} \omega_{X'X} e^{i \pmb{G}_i \cdot \pmb{\tau}_{X'X}}\frac{\prod_{m=1}^{n} \gamma_{i j_m}'}{n!} \exp\left[i \left(\pmb{q}_{\eta} + \pmb{g}_i + \sum_{m=1}^{n} \pmb{g}_{j_m}\right) \cdot \pmb{r}\right],
\end{equation}
\begin{equation}\label{eq:exband_intra}
    V_{X'X}^{l,(n)}(\pmb{r}) = \sum_{i=1}^{6} \sum_{j_1, \dots, j_n=1}^{6} \nu_{X'X}^{l,i} \frac{\prod_{m=1}^{n} \gamma_{i j_m}}{n!} \exp\left[i \left(\pmb{g}_i + \sum_{m=1}^{n} \pmb{g}_{j_m}\right) \cdot \pmb{r}\right].
\end{equation}
Explicitly for low orders
\begin{equation}
    \begin{aligned}
        T^{(0)}_{X'X}(\pmb{r}) &= \sum_{i \in \{0,2,3\}} \omega_{X'X} \exp\left[i \pmb{G}_i \cdot \pmb{\tau}_{X'X}\right] \exp\left[i (\pmb{g}_i + \pmb{q}_{\eta}) \cdot \pmb{r}\right], \\
        T^{(1)}_{X'X}(\pmb{r}) &= \sum_{i \in \{0,2,3\}} \sum_{j=1}^{6} \omega_{X'X} \gamma_{ij}' \exp\left[i \pmb{G}_i \cdot \pmb{\tau}_{X'X}\right] \exp\left[i (\pmb{q}_{\eta} + \pmb{g}_i + \pmb{g}_j) \cdot \pmb{r}\right], \\
        T^{(2)}_{X'X}(\pmb{r}) &= \sum_{i \in \{0,2,3\}} \sum_{j_{1},j_{2}=1}^{6} \omega_{X'X} \frac{\gamma_{ij_{1}}' \gamma_{ij_{2}}'}{2} \exp\left[i \pmb{G}_i \cdot \pmb{\tau}_{X'X}\right] \exp\left[i (\pmb{q}_{\eta} + \pmb{g}_i + \pmb{g}_{j_{1}} + \pmb{g}_{j_{2}}) \cdot \pmb{r}\right], \\
        &\cdots
    \end{aligned}
\end{equation}
\begin{equation}
    \begin{aligned}
        V_{X'X}^{l,(0)}(\pmb{r}) &= \sum_{i=1}^{6} \nu_{X'X}^{l,i} \exp\left[i \pmb{g}_i \cdot \pmb{r}\right], \\
        V_{X'X}^{l,(1)}(\pmb{r}) &= \sum_{i=1}^{6} \sum_{j=1}^{6} \nu_{X'X}^{l,i} \gamma_{ij} \exp\left[i (\pmb{g}_i + \pmb{g}_j) \cdot \pmb{r}\right], \\
        V_{X'X}^{l,(2)}(\pmb{r}) &= \sum_{i=1}^{6} \sum_{j_{1},j_{2}=1}^{6} \nu_{X'X}^{l,i} \frac{\gamma_{ij_{1}} \gamma_{ij_{1}}}{2} \exp\left[i (\pmb{g}_i + \pmb{g}_{j_{1}} + \pmb{g}_{j_{2}}) \cdot \pmb{r}\right]. \\
        &\cdots
    \end{aligned}
\end{equation}

\begin{figure}[t]
    \centering
    \includegraphics[width=0.5\textwidth]{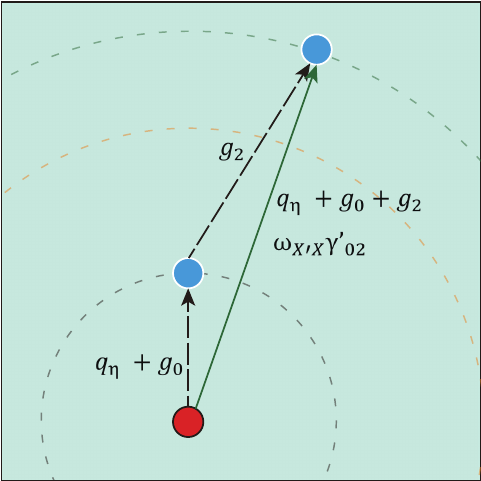}
    \caption{\label{fig:hop}The hopping term corresponding to $i=0, j=2$ in $T^{(1)}_{X'X}(\pmb{r})$. Since $\pmb{G}_{0}=0$, the phase $e^{i \pmb{G}_0 \cdot \pmb{\tau}_{X'X}}=1$.}
\end{figure}

The $n=0$ case corresponds to the rigid case. Terms with $n \geq 1$ provide successive corrections for relaxation effects. 
As a schematic illustration, we show in Fig.~\ref{fig:hop} the hopping term corresponding to $i=0, j=2$ in $T^{(1)}_{X'X}(\pmb{r})$, note that since $\pmb{G}_{0}=0$, the phase $e^{i \pmb{G}_0 \cdot \pmb{\tau}_{X'X}}=1$.
This expansion establishes an explicit connection between lattice relaxation and electronic band reconstruction, incorporating atomic-scale information into the continuum model.

\subsection{ Range of Validity}
The convergence of the expansions Eq.~(\ref{eq:exband_Q}) and Eq.~(\ref{eq:exband_G}) requires $\gamma_{ik}', \gamma_{jk} < 1$. The maximum value is $\gamma_{\text{max}} = \frac{\kappa_{\parallel}}{2} \frac{1}{\theta^{2}}$. Thus, we require
\begin{equation}
    \frac{\kappa_{\parallel}}{2} \frac{1}{\theta^{2}} < 1 \implies \theta > \sqrt{\frac{\kappa_{\parallel}}{2}}.
\end{equation}
For example, in TMDs $\frac{\kappa_{\parallel}}{2} \approx 10^{-4}$ requires $\theta > 1^{\circ}$; in graphene $\frac{\kappa_{\parallel}}{2} \approx 10^{-5}$ requires $\theta > 0.3^{\circ}$. At twice the critical angle, $T^{(1)} \sim 0.25 T^{(0)}$ and $T^{(2)} \sim 0.03125 T^{(0)}$, ensuring rapid convergence.

\subsection{Many-Body Calculation Details}

In our many-body calculations for the twisted MoTe$_{2}$ system, the fully spin (or valley) polarization is assumed at hole filling $\nu=\frac{2}{3}$\cite{cai_signatures_2023,zengThermodynamicEvidenceFractional2023,xuObservationIntegerFractional2023}. The Hamiltonian can be written as $H = H_{0} + H_{\text{int}}$. Considering hole doping, the single-particle part is given by
\begin{equation}
H_{0} = -\sum_{\pmb{k}} \epsilon_{\pmb{k}} \gamma^{\dagger}_{\pmb{k}} \gamma_{\pmb{k}}.
\end{equation}
The interaction part is expressed as
\begin{equation}
H_{\text{int}} = \frac{1}{2A} \sum_{\pmb{k}\pmb{k}^{'}\pmb{q},ll^{'}} V(\pmb{q})c^{\dagger}_{\pmb{k}+\pmb{q},l}c^{\dagger}_{\pmb{k}^{'}-\pmb{q},l^{'}}c_{\pmb{k}^{'},l^{'}}c_{\pmb{k},l},
\end{equation}
where $V(\pmb{q}) = \frac{2\pi e^{2}}{\epsilon |\pmb{q}|}$ and $A$ denotes the total area of the system, and $c_{\pmb{k}}$ is the electron annihilation operator. By diagonalizing the single-particle Hamiltonian via $\gamma^{\dagger}_{\pmb{k}} = \sum_{\pmb{G}, l} u_{\pmb{G}, l}(\pmb{k}) c^{\dagger}_{\pmb{k} + \pmb{G}, l}$, we project the interactions onto the top band
\begin{equation}
\begin{aligned}
H_{\text{int}} &= \frac{1}{2A} \sum_{\{\pmb{k}_{i}\}, \pmb{q}} V^{\pmb{k}_1 \pmb{k}_3}_{\pmb{k}_2 \pmb{k}_4}(\pmb{q}) \gamma^{\dagger}_{\pmb{k}_{1}} \gamma^{\dagger}_{\pmb{k}_{2}} \gamma_{\pmb{k}_3} \gamma_{\pmb{k}_4} \\
&= \frac{1}{2A} \sum_{\{\pmb{k}_{i}\}, \pmb{q}} \frac{2\pi e^{2}}{\epsilon |\pmb{q}|} F(\pmb{k}_{1}, \pmb{k}_{3}, \pmb{G}_{\pmb{k}_{3}+\pmb{q}}) F(\pmb{k}_{2}, \pmb{k}_{4}, \pmb{G}_{\pmb{k}_{4}-\pmb{q}}) \gamma^{\dagger}_{\pmb{k}_{1}} \gamma^{\dagger}_{\pmb{k}_{2}} \gamma_{\pmb{k}_3} \gamma_{\pmb{k}_4},
\end{aligned}
\end{equation}
where
\begin{equation}
    F(\pmb{k},\pmb{k}^{'},\pmb{G}_{\pmb{k}})=\sum_{\pmb{G}_{\pmb{k}^{'}},l}u^{*}_{\pmb{G}_{\pmb{k}^{'}}+\pmb{G}_{\pmb{k}},l}(\pmb{k})u_{\pmb{G}_{\pmb{k}^{'}},l}(\pmb{k}^{'}).
\end{equation}
Here, $\pmb{G}_{\pmb{k}+\pmb{q}}$ denotes the reciprocal lattice vector that folds $\pmb{k}+\pmb{q}$ back into the first moiré Brillouin zone. In the exact diagonalization calculations, we adopt a cluster configuration with $N =N_{1}\times N_{2}= 9 \times 3 = 27$,
here $N_i$ corresponds to the number of supercell repetitions along the moiré lattice vectors $\bm{a}^{M}_{i}$\cite{regnault_fractional_2011}. We define the center-of-mass momentum at each point as $\bm{K}_{\text{com}} = k_1 \bm{T}_1 + k_2 \bm{T}_2$, and label it via $\bm{K}_{\text{com}} = k_1 N_2 + k_2$ (where $k_i = 0, \cdots, N_i - 1$). Here, $\bm{T}_i$ denotes the reciprocal lattice vectors of the supercell.

The many-body Chern number is calculated by imposing twisted boundary conditions on the many-body wave function, introducing twist phases $\theta_1$ and $\theta_2$ that play the role of magnetic fluxes. The Chern number is defined as\cite{sheng_fractional_2011,niu_quantized_1985}
\begin{equation}
\mathcal{C}_\mathrm{MB} = \frac{1}{2\pi} \int_0^{2\pi} d\theta_1 \int_0^{2\pi} d\theta_2 \mathrm{Im}\left( \left\langle \frac{\partial \Psi(\pmb{\theta})}{\partial \theta_2} \middle| \frac{\partial \Psi(\pmb{\theta})}{\partial \theta_1} \right\rangle - \mathrm{c.c.} \right).
\end{equation}
We can define
\begin{equation}
A_i(\pmb{\theta}) = \frac{\langle \Psi(\pmb{\theta}) | \Psi(\pmb{\theta} + \delta\theta_i) \rangle}{\left| \langle \Psi(\pmb{\theta}) | \Psi(\pmb{\theta} + \delta\theta_i) \rangle \right|},
\end{equation}
and the many-body Berry curvature on a small plaquette is\cite{fukuiChernNumbersDiscretized2005}
\begin{equation}
\Omega_{12}(\pmb{\theta})=\mathrm{ln}A_{1}(\pmb{\theta})A_{2}(\pmb{\theta}+\delta\theta_{1})A_{1}^{-1}(\pmb{\theta}+\delta\theta_{2})A_{2}^{-1}(\pmb{\theta}).
\end{equation}
The many-body Chern number is then obtained by summing the imaginary part of $\Omega_{12}$ over the entire twist-phase space
\begin{equation}
\mathcal{C}_\mathrm{MB}=\frac{1}{2\pi}\int_{0}^{2\pi}d\theta_{1}\int_{0}^{2\pi}d\theta_{2}\mathrm{Im}\,\Omega_{12}(\pmb{\theta}).
\end{equation}
Employing the above method, we investigate twisted MoTe$_2$ at a hole filling of $\nu = \frac{2}{3}$ and the twist angle of $\theta = 3.0^\circ$, including relaxation effects. For the 27-site system, the many-body Chern numbers for the three degenerate states are (1, 1, 0), indicating the formation of a fractional Chern insulator state under these conditions.

\subsection{Results}
Having analytically established the effects of in-plane and out-of-plane relaxation on twisted electronic structures, we now apply this framework to two prominent systems: tTMD and TBG.

For tTMD, we employ the continuum model parameters from Ref.\cite{wu_topological_2019}. Crucially, their DFT parameterization for untwisted bilayers incorporates out-of-plane relaxation corrections. However, in-plane relaxation—being periodic-direction-dependent—cannot be captured by simple untwisted DFT calculations. For clarity, we retain the term "rigid" for the n=0 case (no in-plane relaxation).

\begin{figure}[t]
    \centering
    \includegraphics[width=1.0\textwidth]{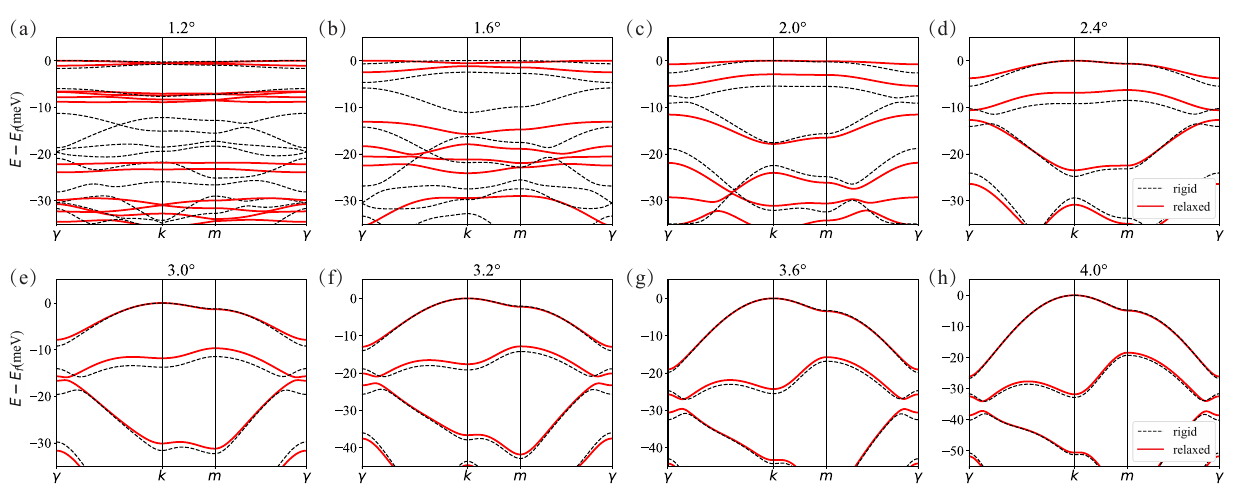}
    \caption{\label{fig:tTMD_band}Band structures of tTMD under relaxation compared with the rigid case, the twist angle from $\theta=1.2^{\circ}$ to $\theta=4.0^{\circ}$.}
\end{figure}

\begin{figure}[t]
    \centering
    \includegraphics[width=0.8\textwidth]{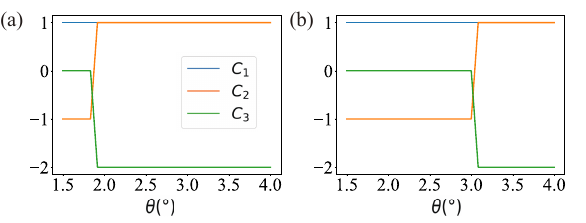}
    \caption{\label{fig:tTMD_topo} Evolution of topological structure for the first three bands in tMoTe$_{2}$ with twist angle $\theta$. (a) $n=0$: pure rigid case, phase transition occurs at $\sim 1.8^{\circ}$. (b) $n=2$: including first-order in-plane relaxation effects, phase transition shifts to $\sim 3^{\circ}$.}
\end{figure}

Figure \ref{fig:tTMD_band} compares band structures under rigid (n=0) and relaxed (n=1) conditions at various twist angles. Key findings: At small angles, in-plane relaxation substantially modifies electronic structures; Its influence diminishes with increasing $\theta$, converging toward rigid-limit behavior. Our theory is valid for tTMD at $\theta > 0.8^{\circ}$. Figure \ref{fig:tTMD_topo} reveals a pivotal topological evolution: The rigid model places the topological phase transition at $\theta=1.8^{\circ}$, deviating significantly from experiments and DFT. Including in-plane relaxation shifts the transition to $\theta=3.0^{\circ}$, achieving quantitative agreement with experiments and DFT.

\begin{figure}
    \centering
    \includegraphics[width=0.8\textwidth]{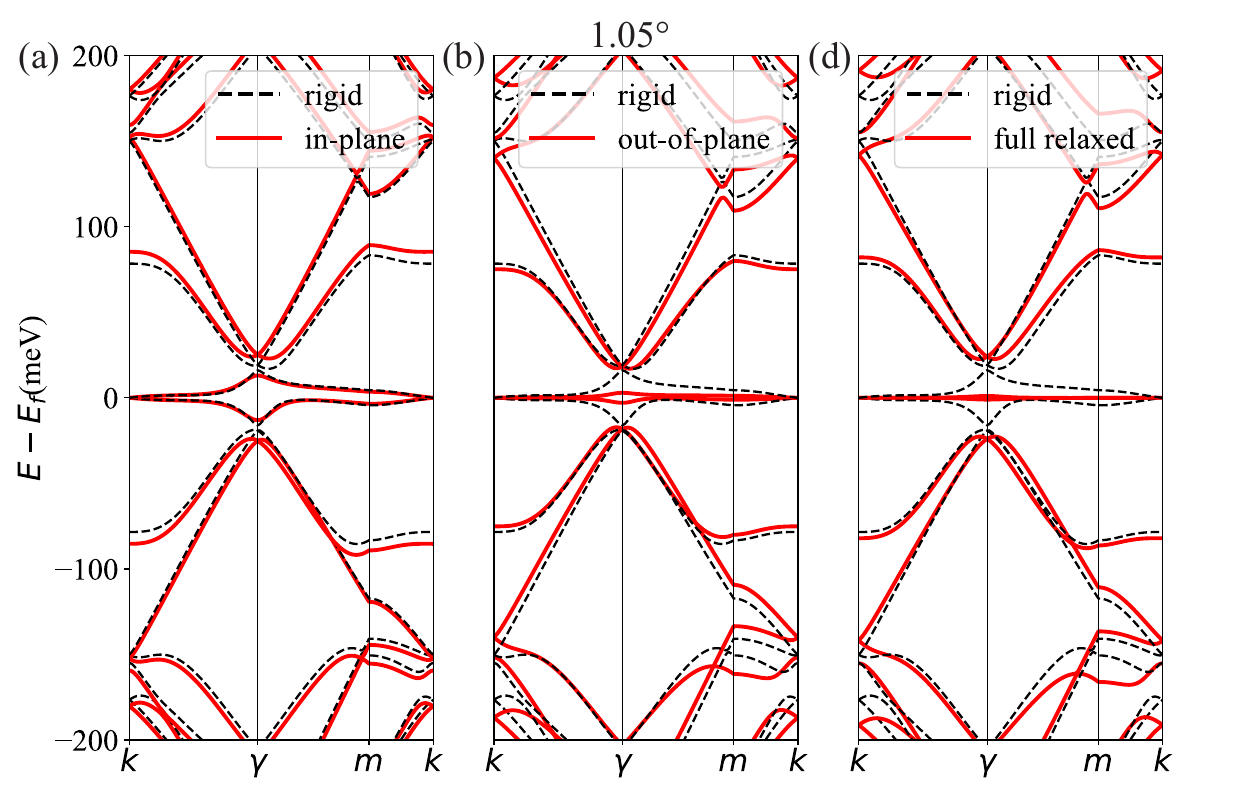}
    \caption{\label{fig:TBG_band_105}Band structures of TBG under distinct relaxation conditions compared with the rigid case, at twist angle $\theta=1.05^{\circ}$. (a) Considering only in-plane relaxation. (b) Considering only out-of-plane relaxation. (c) Full relaxation.}
\end{figure}

\begin{figure}
    \centering
    \includegraphics[width=0.8\textwidth]{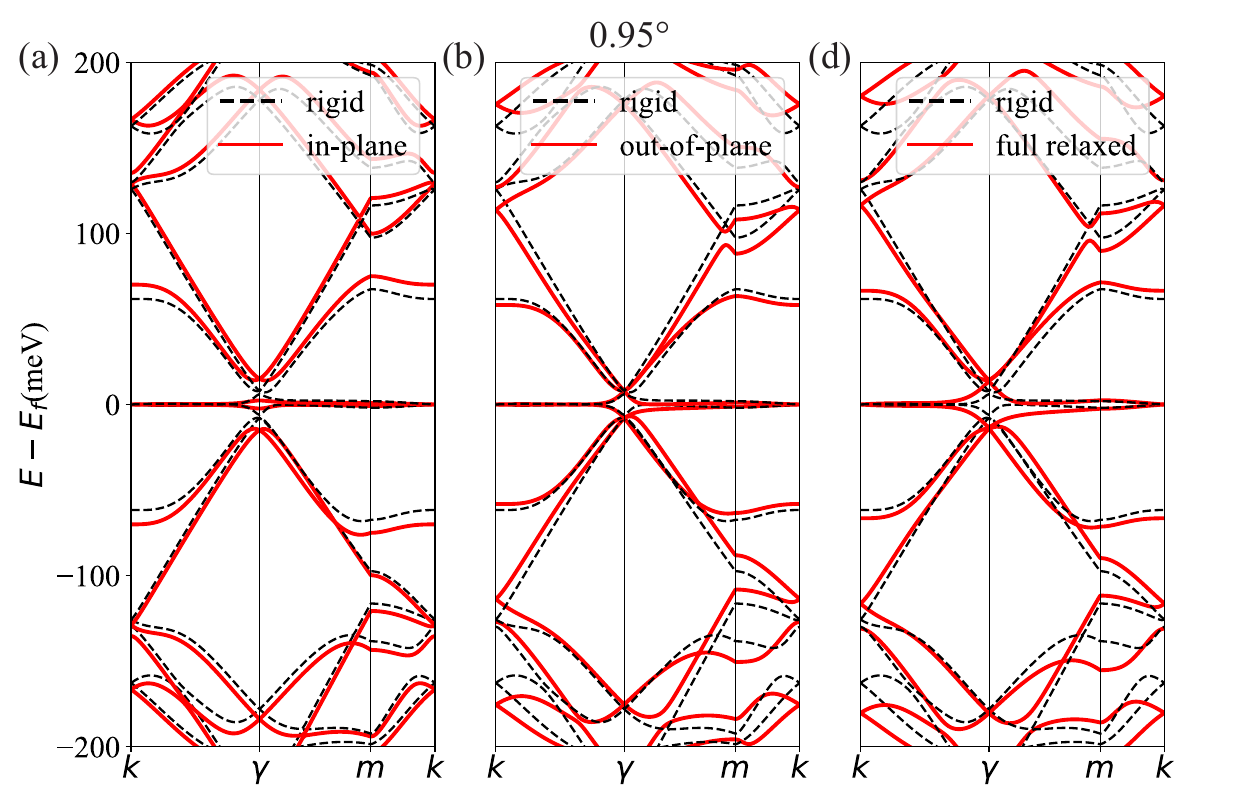}
    \caption{\label{fig:TBG_band_095}Band structures of TBG under distinct relaxation conditions compared with the rigid case, at twist angle $\theta=0.95^{\circ}$. (a) Considering only in-plane relaxation. (b) Considering only out-of-plane relaxation. (c) Full relaxation.}
\end{figure}

\begin{figure}
    \centering
    \includegraphics[width=0.8\textwidth]{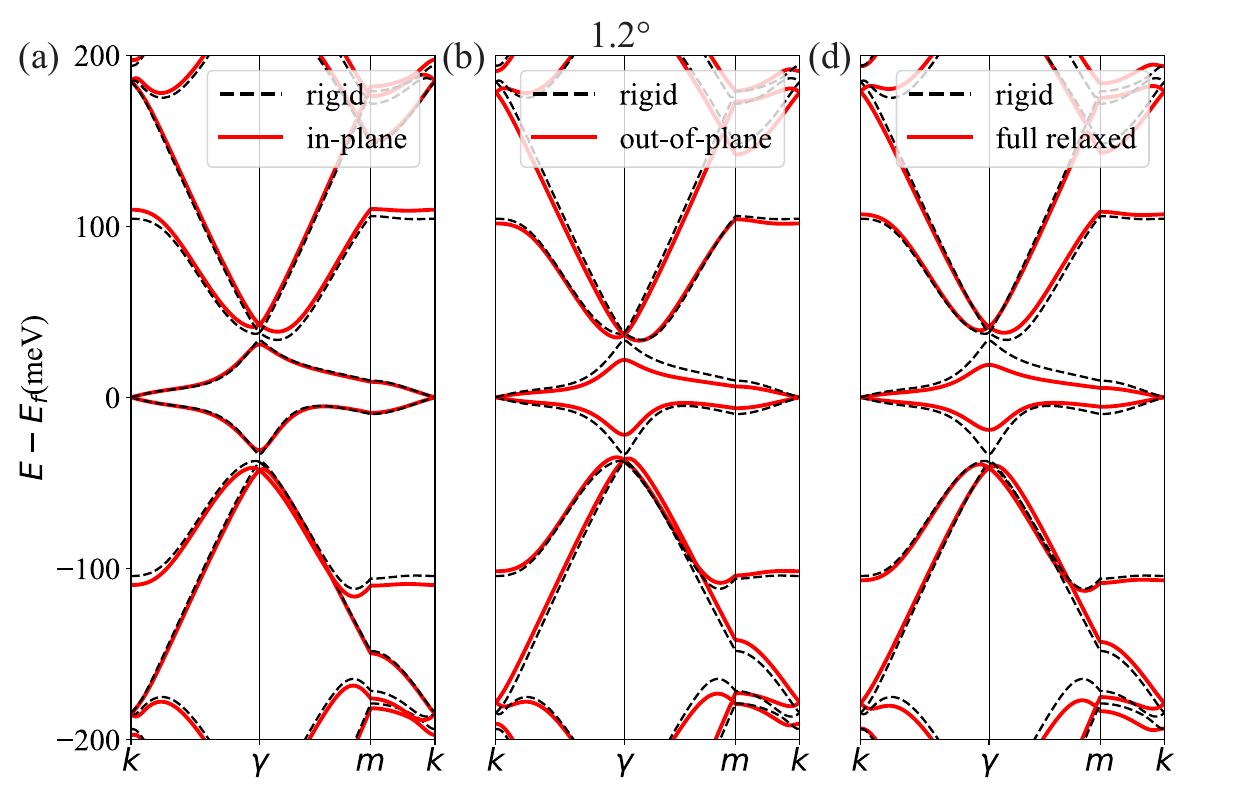}
    \caption{\label{fig:TBG_band_120}Band structures of TBG under distinct relaxation conditions compared with the rigid case, at twist angle $\theta=1.2^{\circ}$. (a) Considering only in-plane relaxation. (b) Considering only out-of-plane relaxation. (c) Full relaxation.}
\end{figure}

\begin{figure}
    \centering
    \includegraphics[width=0.8\textwidth]{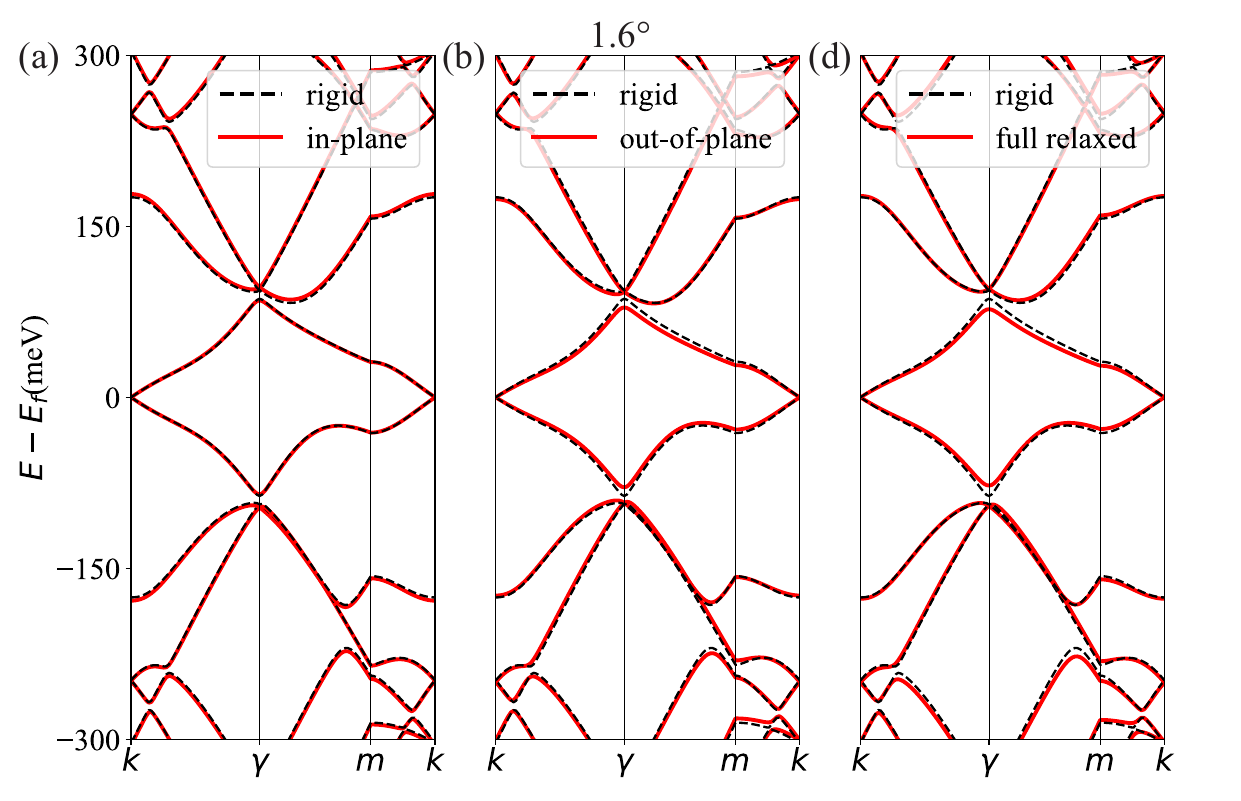}
    \caption{\label{fig:TBG_band_160}Band structures of TBG under distinct relaxation conditions compared with the rigid case, at twist angle $\theta=1.6^{\circ}$. (a) Considering only in-plane relaxation. (b) Considering only out-of-plane relaxation. (c) Full relaxation.}
\end{figure}

\begin{figure}
    \centering
    \includegraphics[width=0.8\textwidth]{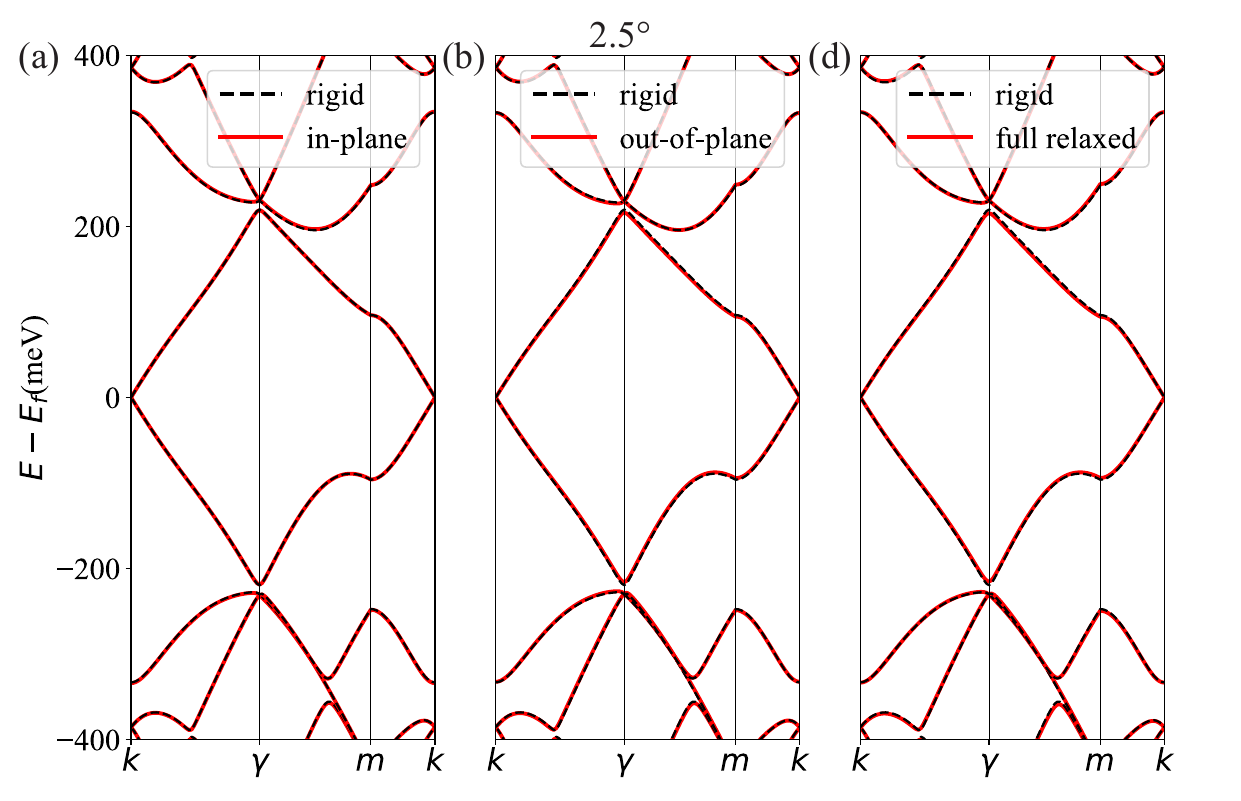}
    \caption{\label{fig:TBG_band_250}Band structures of TBG under distinct relaxation conditions compared with the rigid case, at twist angle $\theta=2.5^{\circ}$. (a) Considering only in-plane relaxation. (b) Considering only out-of-plane relaxation. (c) Full relaxation.}
\end{figure}

Applying the framework to TBG, Fig. \ref{fig:TBG_band_105} demonstrates that near the magic angle ($\theta=1.05^{\circ}$): (i) Rigid bands exhibit degeneracy between low-energy and excited states; (ii) Introducing either relaxation mechanism lifts this degeneracy and reduces low-energy bandwidth; (iii) Full relaxation yields nearly exactly flat bands.
This is explained by stacking-region contribution imbalance: In-plane relaxation shrinks AA domains while expanding AB regions; Out-of-plane relaxation increases AA interlayer spacing (weakening coupling) while decreasing AB spacing (enhancing coupling). Both effects suppress AA-region contributions, consistent with DFT and experiments \cite{cantele_structural_2020,kazmierczak_strain_2021}. Notably, out-of-plane relaxation exerts a more pronounced effect in TBG due to the inherently lower out-of-plane stiffness of graphene. TBG with several other twist angles are shown in Figs.~\ref{fig:TBG_band_095}-\ref{fig:TBG_band_250}.

Our framework thus quantifies relaxation effects across twisted systems and elucidates their physical origins.


\bibliography{SMref}